\documentclass[10pt, journal, compsoc]{IEEEtran}
\IEEEoverridecommandlockouts

\usepackage{tikz}
\usetikzlibrary{calc,positioning,arrows}
\usetikzlibrary{positioning,fit,calc}
\usetikzlibrary{automata,shapes}
\usepackage{listings}
\graphicspath{ {./img/} }

\usepackage{cite}
\usepackage{amsmath,amssymb,amsfonts}
\usepackage[ruled, lined, linesnumbered]{algorithm2e}
\usepackage{graphicx}
\usepackage{textcomp}
\usepackage{xcolor}
\usepackage{tabularx}
\usepackage{xspace}
\usepackage{cleveref}
\usepackage{csquotes}
\usepackage{caption}
\usepackage{subcaption}
\usepackage{url}
\usepackage{dblfloatfix} 
\usepackage{balance}
\usepackage{color}

\definecolor{gray}{rgb}{0.4,0.4,0.4}
\definecolor{darkblue}{rgb}{0.0,0.0,0.6}
\definecolor{cyan}{rgb}{0.0,0.6,0.6}

\crefformat{section}{\S#2#1#3}
\crefformat{subsection}{\S#2#1#3}
\crefformat{subsubsection}{\S#2#1#3}
\crefrangeformat{section}{\S\S#3#1#4 to~#5#2#6}
\crefmultiformat{section}{\S\S#2#1#3}{ and~#2#1#3}{, #2#1#3}{ and~#2#1#3}
\crefname{listing}{Schema}{Schemes}

\lstdefinestyle{mycode}{
    basicstyle=\scriptsize,
    stringstyle=\ttfamily,
    breakatwhitespace=false,         
    breaklines=true,                 
    captionpos=b,                    
    keepspaces=true,                                                      
    showspaces=false,                
    showstringspaces=false,
    showtabs=false,                  
    tabsize=1,
    frame=tb,
    float=t,
}

\lstdefinestyle{schema}{%
  frame=tb, float=b,
}

\lstdefinelanguage{mylang}
{
  identifierstyle=\color{darkblue},
  stringstyle=\color{black},
  keywordstyle=\color{cyan},
  morestring=[s]{\{t}{\}},
}

\newcommand{\sys}{\texttt{Oakestra}\xspace}

\newcommand*\circled[1]{\tikz[baseline=(char.base)]{
            \node[shape=circle,draw,inner sep=0.6pt] (char) {#1};}}
            
\newcommand{\linebreakand}{%
  \end{@IEEEauthorhalign}
  \hfill\mbox{}\par
  \mbox{}\hfill\begin{@IEEEauthorhalign}
}
\def\BibTeX{{\rm B\kern-.05em{\sc i\kern-.025em b}\kern-.08em
    T\kern-.1667em\lower.7ex\hbox{E}\kern-.125emX}}

\newcommand{\hrulealg}[0]{\vspace{1mm} \hrule \vspace{1mm}}

\SetCommentSty{mycommfont}

\begin{document}

\title{Oakestra white paper: \\ {An Orchestrator for Edge Computing}}


\author{

%
%
%
%
%


Giovanni~Bartolomeo,~Mehdi~Yosofie,~Simon~Bäurle,~Oliver~Haluszczynski,~Nitinder~Mohan~and~Jörg~Ott

}

\bstctlcite{IEEEexample:BSTcontrol}

\IEEEtitleabstractindextext{%
\begin{abstract}

Edge computing seeks to enable applications with strict latency requirements by utilizing compute resources deployed closer to the users. The diverse, dynamic, and constrained nature of edge infrastructures necessitates a flexible orchestration framework that dynamically supports application QoS requirements. However, existing state-of-the-art orchestration platforms were designed for datacenter environments and make strict assumptions about underlying infrastructures that do not hold for edge computing. This work proposes a novel hierarchical orchestration framework specifically designed for supporting service operation over edge infrastructures. Through its novel federated cluster management, delegated task scheduling, and semantic overlay networking, our system can flexibly consolidate multiple infrastructure operators and absorb dynamic variations at the edge. We comprehensively evaluate our proof-of-concept implementation -- \sys\ -- against state-of-the-art solutions in both controlled and realistic testbeds and demonstrate the significant benefits of our approach as we achieve $\approx$ 10$\times$ and 30\% reduction in CPU and memory consumption, respectively.

\end{abstract}

}

\maketitle

\IEEEpeerreviewmaketitle



\IEEEraisesectionheading{\section{Introduction}\label{sec:introduction}}

%
\IEEEPARstart{W}{ithin} almost a decade since its inception, edge computing has found a wide range of use cases in both industry and research, especially for supporting sophisticated services like AR/VR, live video analytics etc.~\cite{mohan2020pruning, edgeai, nimbus}.
However, despite significant interest in the field, there have only been a handful of real-world demonstrations of the paradigm so far~\cite{noghabi2020emerging}.
We attribute the cause to the following reasons.
%
%
%
%
Firstly, the resource capacity significantly decreases as we move out of the datacenter silos and towards the network's edge to match the close-proximity requirements~\cite{promise_of_edge} and smaller form factor specialized hardware with CPU, GPU, TPU and VPU become more prevalent~\cite{intelnuc:online, coral:online}.
%
However, unlike siloed deployment of cloud datacenters, setting up edge infrastructures requires significant investment and planning as the compute fabric co-exists alongside its users, e.g., in base stations, city-owned properties, public transport, etc.~\cite{noghabi2020emerging, efcloud, decloud, mute, placing2018mohan}.
Furthermore, the benefits of edge computing are only apparent when there is a dense availability of computing resources~\cite{mohan2020pruning, surrounded, cloudy, howmuch}.
%

Secondly, the majority of de-facto resource orchestration and application deployment frameworks, e.g., Kubernetes~\cite{Kubernetes:online}, K3s~\cite{k3s:online}, KubeFed~\cite{kubefed}, etc., are either off-shoot branches or adoptions of datacenter-oriented solutions and, therefore, struggle to leverage the edge. 
Specifically, such approaches
make strong assumptions regarding the underlying infrastructure's consistent reliability and reachability, which does not hold for edge computing.
Similarly, finding optimal service deployment in a loosely coupled infrastructure spanning vast geographical regions has been proven to be a non-trivial problem~\cite{7541539,7870615,fog-placement}, which is almost entirely overlooked by such works.
Furthermore, almost none of the existing scheduling platforms can currently support the significant heterogeneity and diversity in processing, 
networking~\cite{cloudy, qaware}, 
and availability of resources~\cite{anveshak}, which are synonymous with edge computing.

This paper presents a flexible hierarchical orchestration framework that overcomes the many challenges of edge infrastructures and workloads.
%
Our system allows multiple operators to contribute their resources to shared infrastructure and retain administrative control -- thereby significantly reducing the effort to achieve a dense computing fabric at the edge.
The key innovation lies in the edge-focused design of our system components that allows the framework to cope with infrastructure complexity while providing application developers familiar experience for deploying and managing their services.
Specifically, we make the following contributions:
%

\smallskip
\noindent
\textbf{(1)}  We propose a \emph{hierarchical} resource orchestration scheme that decomposes the control-plane management across segregated tiers of clusters.
Each resource is managed by its local cluster orchestrator which coordinates with its parent orchestrator for exchanging aggregated statistics and deployment commands. 
Our approach allows flexible infrastructure scaling at the edge by providing context separation between resources managed by different cluster operators (\cref{section:system_overview}). 

\smallskip
\noindent
\textbf{(2)} We propose a \emph{delegated service scheduling mechanism} that decentralizes the task placement problem across the hierarchy to effectively support service deployment at scale (\cref{section:scheduling}). 
Similar to cloud systems, application developers can define high-level service operational requirements using our SLA definition.
%
Our system offloads the SLA to best-candidate cluster orchestrators, which then find a suitable resource for supporting the service within their operational boundaries.
We also present a novel \emph{latency and distance placement} (LDP) algorithm that optimizes latency and geographical distance constraints while deploying services at the edge.
%
%
%
%
%

\smallskip
\noindent
\textbf{(3)} We design a robust overlay network that enables service interactions across edge resources in different (private) networks without overheads (\cref{section:communication}). 
%
Through our novel \emph{semantic addressing} scheme, we can dynamically (and transparently) adjust communication endpoints in response to infrastructure changes, e.g., service migrations, resource failures, etc., ensuring uninterrupted service interactions.
Our networking component also supports edge-oriented load balancing policies, e.g., connecting to the closest instance, effectively utilizing the geographically vast and diverse edge computing infrastructures.
%
%
%

\smallskip
\noindent
\textbf{(4)} We implement \sys, which is lightweight and features compatibility with technologies popularly used in modern cloud applications (\cref{section:implementation}).
%
Our extensive evaluation conducted in both high-performance computing and realistic edge-like infrastructures shows that \sys consistently outperforms the state-of-the-art by a large margin and efficiently integrates heterogeneous resources (\cref{section:evaluation}).
%
Our experiments using realistic edge workloads (e.g., live video analytics) highlight the effectiveness of \sys for supporting distributed microservice-based applications on edge infrastructures. 
Our results show up to 10$\times$ lower CPU overhead and 60\% reduction in service deployment time.
Under heavy loads, our platform reduces resource utilization by $\approx$ 20\% than its closest competitor.

\section{Background and Related Work}
\label{section:background}

Most available service scheduling and monitoring frameworks were designed for datacenter environments.
%
Of these, Kubernetes~\cite{Kubernetes:online} is the most popular orchestration system in production, used by $\approx$ 59\% large organizations~\cite{KubernetesUse}, and has been touted by many as the primary solution for managing edge infrastructures.
However, Kubernetes' inherent operation makes strong assumptions about the underlying infrastructure, which was found to be its primary limitation when ported to the edge~\cite{rearchitecting-kubernetes-for-the-edge}.
Specifically, the platform requires all resources to be in the same cluster and directly reachable (similar to datacenters) -- requiring a close coupling between the orchestrator and workers. 
%
%
%
Additionally, the default service scheduling policies in Kubernetes are not suited for heterogeneous and diverse edge infrastructures as they do not consider metrics such as end-to-end latency, geographical locations, etc.
%
%
Other frameworks, such as KubeEdge~\cite{kubeedge}, K3s~\cite{k3s:online}, Microk8s~\cite{mk8s:online}, and KubeFed~\cite{kubefed}, have re-architected Kubernetes to make it lightweight and suitable for edge computing.
However, recent explorations have also found these to be restrictive, partly since they inherit the strong infrastructure assumptions of Kubernetes in their design as well~\cite{bohm2021profiling}.
On the other hand, our proposed hierarchical orchestration framework is designed from the ground-up to support edge computing infrastructures and workloads through constraint-aware delegated scheduling (\Cref{section:scheduling}) and semantic overlay networks (\Cref{section:communication}).

Only a few works have explored the effective orchestration of edge servers from a research standpoint. 
CloudPath~\cite{cloudpath} envisions a multi-tier on-path computing paradigm that allows stateless functions to be deployed closer to end-user and IoT devices. 
%
%
%
Projects like HeteroEdge~\cite{heteroEdge} or SpanEdge~\cite{spanEdge} cater specifically to support streaming-based applications on edge servers while FogLamp~\cite{foglamp:online} focuses on data management at the edge. 
VirtualEdge~\cite{virtualedge} aims at supporting orchestration at the edge but is limited to cellular networks.

Researchers have also proposed hierarchical scheduling solutions for the edge similar to ours~\cite{overloaded-edge-sched, lambda-sched, ad-hoc-sched}.
Most of these approaches exploit different domain knowledge by distributing the scheduling across the cloud-to-edge hierarchy. 
In \cite{reduce-traffic-sched}, the authors present an autonomous hierarchical scheduling approach that distributes service tasks on a cloud-to-edge continuum.
However, while the focus of these works is to design a service scheduling solution for the edge, we provide an orchestration solution that offers both service and resource management for edge infrastructures.  
Additionally, as we show in \cref{section:evaluation}, \sys significantly outperforms production orchestration frameworks, 
thereby demonstrating its suitability for edge computing. 
\section{Framework Overview}
\label{section:system_overview}



\begin{figure*}[!ht]
\centering
\includegraphics[width=\textwidth]{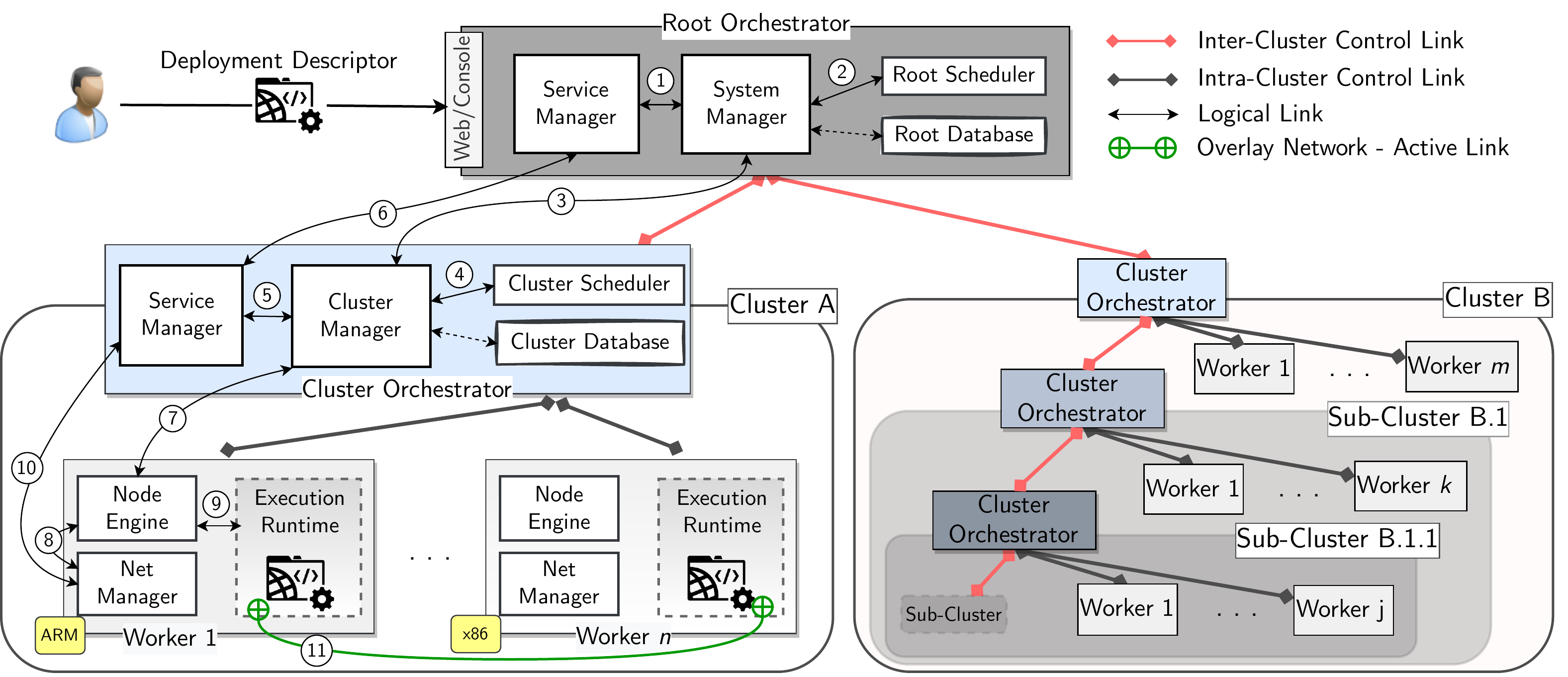}
\caption{System Architecture and Workflow.}
\label{fig:edgeio-schema}
\end{figure*}

\subsection{Challenges \& Design Motivations} \label{subsec:designchallenges}

Previous research has shown that both service deployment and resource management in distributed edge infrastructures are non-trivial problems, primarily due to the heterogeneity and dynamicity of the environment, which convolutes with increasing scale~\cite{7541539,7870615,fog-placement}.
Simultaneously, since the coverage area and capacity of edge servers is significantly smaller than cloud datacenters, application providers are likely to deploy multiple service instances with specialized operational requirements to maximize their client's quality of experience (QoE)~\cite{salaht2020overview}.
As a result, orchestration platforms designed to 
support edge computing
must propose solutions to two significant challenges.
%
\emph{Firstly}, 
the platforms should not only incorporate the heterogeneous edge hardware but also 
support a consolidated, shared infrastructure that allows (a) application developers to utilize edge resources 
regardless of their ownership, and (b) all participating operators to retain complete contextual and management control over their resources~\cite{decentralizingEdge, exec, decloud}.
%
\emph{Secondly},
%
%
%
%
%
the framework must adapt to dynamic infrastructure (and environment) changes without significantly affecting already operational applications.  


%
%
%

\subsection{System Architecture} \label{subsec:sysArch}

We propose a hierarchical orchestration framework for enabling edge computing applications over heterogeneous edge resources.
%
%
%
Through our system's unique multi-cluster resource management, multiple edge operators (e.g., ISPs, cloud operators, city administration, private players, etc.) can contribute their local deployments towards a shared infrastructure while retaining administrative control~\cite{openedge}.
As a result, our framework provides a mechanism to realize a dense edge computing fabric without requiring significant deployment investments.
%
Simultaneously, we allow application providers to seamlessly migrate their services to the edge by specifying high-level SLAs. 
%
To hide the complexity in edge hardware and infrastructure management from application providers, we design a \emph{delegated service scheduling} approach to effectively handle the scale, density, and heterogeneity at the edge.
Similarly, our semantic overlay network with in-built traffic tunneling allows application providers to seamlessly utilize edge resources from different operators without additional management overheads.
%

%

\Cref{fig:edgeio-schema} shows the high-level architecture of our orchestration framework.
Instead of the flat master-slave design (inherent to most orchestration solutions~\cite{kubernetes_2021, kubeedge, kubefed, k3s:online, Eclipse_iofog:online}), our framework organizes edge infrastructure into hierarchical \emph{clusters} (see clusters A and B).
Resources within a cluster (\emph{workers}) are owned and administered by the cluster operator.
We leave the definition of \enquote{cluster} purposefully abstract as a single operator can deploy several clusters to segregate its resources, e.g., by geographical regions.
The hierarchical management extends within each cluster with many sub-clusters attached to their respective parents -- forming a tree-like hierarchy (as shown in cluster B).
Such a design allows infrastructure providers to logically separate their resources in more specific zones, which would, in turn, ease future scalability.
For example, an ISP can operate its infrastructure in different cities as clusters and regions in each city as sub-clusters.
We also allow independent providers with under-utilized hardware to participate as a single resource cluster operator in the global infrastructure.
We separate the resource and service management responsibilities into different components such that both operations can be performed independently of each other. 
Specifically, \textbf{system manager} is responsible for resource availability and fault-tolerance while \textbf{service manager} handles application service deployment and lifecycle management.
As shown in the figure, our framework composes of three functional entities -- \emph{root}, \emph{cluster}, and \emph{worker}. We now detail the operation of each component.

\subsubsection{Root Orchestrator}

The \textit{root orchestrator} is the centralized control plane of the framework.
The component is analogous to the \enquote{control-plane} of Kubernetes~\cite{kubernetes_2021} and is responsible for managing participating resource clusters.
We envision the root orchestrator to be deployed in the cloud or a node reachable from all clusters.
Developers interested in deploying their applications at the edge submit the application code and a list of service level agreements (SLA) to the \textbf{service manager} in root orchestrator via an API.
The SLA includes high-level operational requirements and constraints for service execution at the edge, e.g., virtualization technology, hardware capacity, geographical location, etc. (detailed in \Cref{subsec:serviceSch}).
%
The service manager notifies the \textbf{system manager} of the new deployment request (step \circled{1}), which registers the service in the local database. 
The system manager contacts the \textbf{root scheduler} (step \circled{2}) to calculate a priority list of clusters best suited to deploy the application. 
We design task placement as a multi-step mechanism that distributes the scheduling operation across schedulers in root and clusters (see \Cref{section:scheduling}).
The system manager is also responsible for registering new clusters to the platform and coordinating information exchange between the cluster and the root orchestrator. 
%
Once the service is deployed, the service manager monitors its operational requirements, e.g., service addressing, external discovery requests, inter-cluster service-to-service communication, etc., and takes remedial actions in case of violations (see \Cref{section:communication}).   
The database stores the current state of all submitted services and all reported operational information from attached cluster orchestrators~\cite{managingData}.

\subsubsection{Cluster Orchestrator} \label{subsec:clusterOrch}

Component-wise, the cluster orchestrator is a logical twin of the root, but
with management responsibility restricted to resources in the cluster.
%
Any infrastructure provider (e.g., ISP) can register its resources as a consolidated cluster with the root orchestrator via an API.
%
In this case, the operator also assigns the orchestrator role to a node that is ideally reachable by every other resource in the cluster.
The cluster orchestrator includes \textbf{service manager}, \textbf{cluster manager} and \textbf{cluster scheduler} components -- which perform similar operations to their counterparts at the root.
%
The cluster manager periodically updates the root system manager with \emph{aggregated} statistics of cluster utilization and deployed services via the \emph{inter-cluster control link}.
Note that the cluster orchestrator withholds minute information 
of its member resources to retain administrative control within cluster boundary (see \cref{subsec:resMgmt} for resource management details). 
%
Through \textit{delegated service scheduling}, we exploit the logical information separation between root and the cluster and the increased resource reachability within each cluster to minimize the overhead for 
task deployment at the root. 
%
Specifically, while scheduling services in the infrastructure, the root scheduler only calculates a list of candidate clusters 
by matching the service SLA constraints to aggregated statistics of attached clusters.
%
It further \emph{delegates} the precise service scheduling task to filtered \textbf{cluster schedulers} in 
highest-priority-first fashion (step \circled{3} and \circled{4}).
As shown in cluster B, a single cluster can host a multi-tier hierarchy of sub-clusters -- each sub-cluster operating as an independent administrative domain of resources. 
Note that there is no difference between a cluster and a sub-cluster from a system perspective other than the parent resource that acts as the orchestrator.
In the case of a multi-tiered cluster hierarchy, the service scheduling task is iteratively delegated down the branch until a suitable edge server for deploying the service is found (see \cref{section:scheduling} for details).
Each cluster (and sub-cluster) \textbf{service manager} periodically sends the health and QoS of all operational services within its domain to its respective parent (step \circled{5} and \circled{6}).

\subsubsection{Worker Node}

We term edge servers responsible for executing services as \emph{workers} which are also the leaves of our orchestration hierarchy.
%
%
Each worker node has distinct capacity and capability, e.g., CPU cores, local disk size, supported execution runtimes etc., which it reports to the cluster orchestrator at the time of registration.
If a worker is found suitable for the requested service's SLA constraints, the cluster orchestrator instructs the worker's \texttt{NodeEngine} to deploy the service  (step \circled{7}).
%
The worker first reserves the required subnetwork for service communication requirements and instantiates the service inside the execution runtime (step \circled{8} and \circled{9}).
Each worker node periodically reports its detailed utilization metrics along with the health of operational services (e.g., SLA default alarms) to its cluster orchestrator via the \emph{intra-cluster control link}.
%
%
It must be noted that we do \emph{not} require worker machines to have public IP addresses but instead assume that workers within a cluster 
can only directly access resources within the same (and parent) cluster.
In case a deployed service needs to communicate with another service in the system (within same or across clusters), the \texttt{NetManager} in the worker fetches the target IP address from the cluster service manager (step \circled{10}) and establishes the connection via an invisible tunnel (step \circled{11}). We detail the network management in \cref{section:communication}. 
%
%



\section{Resource \& Service Management} \label{section:scheduling}


%
The majority of existing orchestration frameworks, e.g., Kubernetes \cite{Kubernetes:online}, K3s \cite{k3s:online}, etc. were designed for cloud infrastructures and workloads~\cite{requirementsForCloud} and, therefore, follow a flat centralized (master-slave) management architecture.
%
Such platforms require all resources in the infrastructure to be homogeneous and consistently available -- an assumption that does not always hold for edge infrastructures~\cite{not-enough}. 
%
%
%
%
Previous research has shown service placement at the edge to be an NP-hard problem \cite{cui2017cloud,7541539,7870615,fog-placement} which increases in complexity as the variables in the system increase.
%
Consequently, frameworks that inherently rely on a centralized orchestrator 
cannot operate at the edge 
without significant overheads~\cite{7541539}.
%
%
%
%
We overcome these challenges by decentralizing the resource management and service scheduling decisions across cluster tiers.
Our design leverages the increased resource reachability within cluster boundaries and, thereby, reduces the dependencies over inconsistent inter-cluster network links for control decisions.  


\subsection{Resource Management} \label{subsec:resMgmt}

As discussed in \cref{subsec:sysArch}, edge
resources in our system participate in distinct clusters and sub-clusters.
We defined such a hierarchy as an oriented tree $I$ such that $I=\langle C,E \rangle$.
$C$ is the set of the clusters $\{C_i | C_i=\{R^i_1...R^i_n,R^i_{CO} \}; {0<i,j\leq|C|}; i\neq j; R^i_l \neq R^j_m \}\bigcup C_0$, and $C_0=\{RO\}$.
Here, $RO$ denotes the root orchestrator, $R^i_{CO}$ is the cluster orchestrator of the $i$-th cluster and $R^i_l$ is the $l$-th resource of the $i$-th cluster. 
We define the edges 
of this tree 
$ E=E_c \bigcup \{ (C_0,C_i) | \nexists_j (C_j,C_i) \in E_c \}$. Here, 
$E_c$ is the set of oriented edges, $(C_i,C_j)$ denoting a inter-cluster control link, i.e. a sub-cluster relationship between $R^i_{CO}$ and $R^j_{CO}$
(see 
\cref{fig:edgeio-schema}).
%

Each resource $R^i_n$ periodically \emph{pushes} its current hardware utilization ($U^i_n$) and other defining characteristics (e.g. location) to $R^i_{CO}$ with update frequency $\lambda({R^i_n})$ over the intra-cluster link.
By correlating $U^i_n$ to the maximum capacity $C^i_n$ of $R^i_n$ reported at registration, $R^i_{CO}$ can monitor the available capacity of each resource in the cluster (denoted by $A^i_n$).
The frequent \textit{push-based} resource status update retains the time-sensitive communication within the cluster where the network is not assumed to be a significant bottleneck.
This relieves the cluster orchestrator of unnecessary management and communication overheads as the infrastructure grows and the proximity between resources decreases significantly.
%
%
%
$\lambda({R^i_n})$ can be different for each resource and can be adjusted dynamically to balance between \enquote{most-updated} information of $R^i_n$ to network overhead caused by frequent updates. 
For example, a worker may only publish an update 
if its $\Delta$ utilization crosses a threshold; or use age-of-information to dynamically adjust the rate to optimize for the system-at-large~\cite{shreedhar2019age, acp2, agecoexistence, ageincloud, ageofinfo}.
We leave the exploration of such solutions for tuning update frequency to future work. 

Similar to intra-cluster operation, update message exchanges are also push-based over inter-cluster links ($C_i$, $C_j$). 
Each cluster orchestrator periodically sends the distribution of \emph{available} cluster and sub-clusters capacities, i.e. $\cup(A^i) = \langle\sum(A^i), \mu(A^i), \sigma(A^i)\rangle$ where $A^i = \{A^i_1, A^i_2, \ldots, A^i_n\} \bigcup \{ A^j | \exists_j (C_i,C_j) \in E \} $, to the orchestrator of the tier above.
The aggregation allows different infrastructure operators to participate in the federated environment while obscuring the minute details of their resources and network.
Additionally, each operator can freely scale up/down its cluster density without involving the parent (or root) orchestrator.
%
%

\subsection{Service Deployment \& Scheduling} \label{subsec:serviceSch}


Application providers can deploy their services on edge servers by specifying QoS requirements as service level agreements (SLA) at the root orchestrator. 
\cref{lst:sla} shows a high-level SLA description supported by our framework.
In addition to operational requirements already prevalent in cloud environments, such as processing performance, networking requirements, virtualization needs, etc., the schema allows developers to specify edge-specific restrictions, e.g., geographical location, specialized hardware, etc. 
%
Additionally, developers can fine-tune the precision of scheduling heuristics by enforcing \emph{convergence time} and \emph{decision rigidness} metrics. 
Convergence time specifies the maximum allowed time within which the scheduler should find the suitable edge server that supports the SLA requirements of the service, and rigidness defines the sensitivity for re-triggering service scheduling in case the selected resource violates the SLA (due to environment/infrastructure changes).

\renewcommand{\lstlistingname}{Schema}
\begin{lstlisting}[caption={Service Requirement Descriptor.},label={lst:sla}, style=mycode, language=mylang]
constraints: [{
    microservice_id: {type: number},
    properties: [{
        memory: {type: integer},
        vcpus: {type: integer},
        vgpus: {type: integer},
        vtpus: {type: integer},
        bandwidth_in: {type: integer},
        latency: {type: number},
        area: {type: string},
        location: {type: string},
        threshold: {type: number},
        rigidness: {type: number},
        convergence_time: {type: integer}, 
        virtualization: {type: string},
        ... }],
    ...}]
\end{lstlisting}

To support application deployment over vast and highly variable edge infrastructures, we propose a \textit{delegated service scheduling} mechanism.
As shown in \Cref{fig:edgeio-schema}, both root and cluster orchestrators have their separate scheduling components that are responsible for solving a subset of the task placement problem within their respective domains.
Let $S = \{ s_1, s_2, \ldots, s_{|S|}\}$ denote the set of services requested to be deployed by the developers at the root.
Each service $s_p \in S$ can be composed of $n$ individual microservices or \textit{tasks}, i.e. $s_p = \{\tau_{p,1}, \tau_{p,2}, \ldots, \tau_{p,n}\}$ where $\tau_{p,i}$ denotes $i$-th task of $p$-th service.
Each task $\tau_{p,i}$ requires a certain capacity (CPU, GPU, memory), denoted by $Q_{\tau_{p,i}}$.
Other considerations like geographical location or virtualization technology, specified by the developer in the SLA, are also part of $Q_{\tau_{p,i}}$.
The task of the scheduling components (in both root and cluster) is to find a suitable resource in the infrastructure that supports the requirements in $Q_{\tau_{p,i}}$.
%
However, since detailed resource availability and utilization information is restricted within cluster boundaries, service scheduling in a $t$-tier edge infrastructure hierarchy is conducted in $t$ steps.

\begin{algorithm}[t]
\footnotesize
\DontPrintSemicolon
\KwIn{$A_n$\hspace{36.5pt}: Information about worker $n$.\\
\hspace{23pt} $Q_{\tau_{p,i}}$\hspace{27pt}:  Requirements of $i$-th task of $p$-th service.\\
\hspace{23pt} $f(A_n,Q_{\tau_{p,i}})$: Resource selection strategy.\\
}
\KwOut{Best worker $W$ to run $\tau_{p,i}$.}
\hrulealg
\tcp{Resource selection strategy examples:}
\tcp{$f(A_n,Q_{\tau_{p,i}}) = \operatorname*{arg\,max}_n \bigl[ (A^{cpu}_n-Q^{cpu}_{\tau_{p,i}}) + (A^{mem}_n - Q^{mem}_{\tau_{p,i}})$}
\tcp{\hspace{85pt} $\:\land\:Q^{virt}_{\tau_{p,i}}\in A^{virt}_n\bigr]$}
\tcp{$f(A_n,Q_{\tau_{p,i}}) = \operatorname*{first}_n \bigl[Q^{cpu}_{\tau_{p,i}}\leq A^{cpu}_n\:\land\:Q^{mem}_{\tau_{p,i}}\leq A^{mem}_n$}
\tcp{\hspace{73pt} $\:\land\:Q^{virt}_{\tau_{p,i}}\in A^{virt}_n\bigr]$}
$W \gets f(A_n,Q_{\tau_{p,i}})$\;
\Return $W$
\caption{Resource-Only Match}\label{alg:rom-pseudocode}
\end{algorithm}
\begin{algorithm}[t]
\footnotesize
\caption{Latency \& Distance Aware Placement}\label{alg:ldp-pseudocode}
\DontPrintSemicolon
\KwIn{$A_n$\hspace{9pt}: Information about worker $n$.\\
\hspace{23pt} $Q_{\tau_{p,i}}$: Requirements of $i$-th task of $p$-th service.\\}
\KwOut{Best workers $W$ to run $\tau_{p,i}$.}
\hrulealg
$W \gets \{n\in [1,|A|]\:|\: A^{cpu}_n \geq Q^{cpu}_{\tau_{p,i}} \land A^{mem}_n \geq Q^{mem}_{\tau_{p,i}}\:\land$
    \hangindent=6.9\skiptext\hangafter=1
    $Q^{virt}_{\tau_{p,i}}\in A^{virt}_n\}$\;
\If{$|Q^{s2s}_{\tau_{p,i}}| \geq 1$}
{
\For{$Q_j$ in $Q^{s2s}_{\tau_{p,i}}$}
{
    $t \gets Q^{trg}_j$\;
    $W \gets \{n \in W \: | \: dist_{gc}(A^{geo}_n,A^{geo}_t) \leq Q^{geo\_thr}_j\:\land$
    \hangindent=5.75\skiptext\hangafter=1
    \hspace{5pt} $dist_{euc}(A^{viv}_n,A^{viv}_t) \leq Q^{viv\_thr}_j\}$\;
}
}
\If{$|Q^{s2u}_{\tau_{p,i}}| \geq 1$}
{
\For{$Q_k$ in $Q^{s2u}_{\tau_{p,i}}$}
{
    $u \gets Q^{lat\_trg}_k$\;
    $rtts \gets \{rtt_{i,u}\:|\: i \in rnd(W), rtt_{i,u} = ping(i,u)\}$\;
    $vivaldiNet \gets \{A^{viv}_n\:|\: n \in [1,|A|]\}$\;
    $\Tilde{U} \gets trilateration(rtts,vivaldiNet)$\;
    $W \gets \{n\in W\:|\: dist_{gc}(A^{geo}_n,Q^{geo\_trg}_k)\leq Q^{geo\_thr}_k\:\land$
    \hangindent=5.75\skiptext\hangafter=1
    $dist_{euc}(A^{viv}_n,\Tilde{U})\leq Q^{lat\_thr}_k$\}\;
}
}
\Return $W$\;
\end{algorithm}

In the first step, the root scheduler matches $Q_{\tau_{p,i}}$ to $\cup(A^i)$ for each cluster such that $\exists_i (C_0,C_i) \in E$ and calculates a priority list of best-fit clusters.
This step filters out all clusters not suitable for the task, e.g., insufficient resource availability, not within target geographical region, no support for the desired virtualization, etc.
The root scheduler then offloads the deployment request to the orchestrator of the cluster with the highest priority.
The request includes both the task $\tau_{p,i}$ and its requirements $Q_{\tau_{p,i}}$.
In the following $t-1$ steps, the respective cluster schedulers either find a suitable worker for the deployment, resulting in early termination of the $t$-step scheduling process, or in turn calculate another best-fit sub-cluster and propagate the deployment request down the branch of the tree $I$.
%
%
Each cluster scheduler can utilize different task placement algorithms to find suitable workers within its boundary depending on the metrics to be optimized~\cite{salaht2020overview}.
In this work, we propose and incorporate two different scheduling approaches.

\smallskip
\noindent \textbf{(1) Resource-Only Match (ROM)}:
As the name suggests, in ROM, the cluster scheduler finds a suitable resource within the cluster that satisfies the capacity requirements of the service (see \Cref{alg:rom-pseudocode}).
%
The scheduling approach is analogous to greedy-fit and knapsack-based solutions popularly used for placing VMs on cloud servers in datacenters~\cite{silva2018approaches}.
%

\smallskip
\noindent \textbf{(2) Latency \& Distance Aware Placement (LDP)}: 
LDP (shown in \Cref{alg:ldp-pseudocode}) builds on the ROM scheduler but additionally considers latency and geographical distance constraints for service placement.
%
%
%
Since edge applications can be composed of multiple microservices that can either interact amongst each other (in a chain-like fashion) or directly with end users/devices, we allow the application provider to specify constraints for both service-to-service (S2S) and service-to-user (S2U) links. 
%
The root scheduler first filters unsuitable clusters by comparing their resource constraints along with approximate geographical operation zones to the SLA requirements.
Within each cluster, 
the algorithm first creates a list of candidate workers that satisfy the resource constraints. 
Then, for all S2S constraints 
$Q^{s2s}_{\tau_{p,i}}$, 
the algorithm filters out workers that exceed the specified distance $Q^{geo\_thr}_j$ and latency thresholds $Q^{viv\_thr}_j$ to the target service $t=Q^{trg}_j$. 
LDP estimates geographic distance as the great circle distance ($dist_{gc}$) between the geographic location of worker $n$ ($A^{geo}_n$) and the location of the target service $A^{geo}_t$. 
The approximated latency is the Euclidean distance ($dist_{euc}$) between the location of worker $n$ ($A^{viv}_n$) and the location of the target service $A^{viv}_t$ in the Vivaldi network~\cite{vivaldi_ncs}. 
Vivaldi is a network coordinate system embedding networked nodes into a $d$-dimensional coordinate system such that the Euclidean distance of two nodes approximates their round-trip time. 
%
If the developer has specified any S2U constraints $Q^{s2u}_{\tau_{p,i}}$, LDP 
measures the round-trip times ($rtts$) to the target as $Q^{lat\_trg}_k$ from a set of random workers in the cluster ($i\in rnd(W)$).
The measurements approximate the user's position within the Vivaldi network via trilateration~\cite{edisco}. 
Following that, LDP filters out workers that either exceed the distance threshold $Q^{geo\_thr}_k$ to
$Q^{geo\_trg}_k$ or the latency threshold $Q^{lat\_thr}_k$ to the approximated user position $\Tilde{U}$.

Suppose the cluster scheduler cannot find appropriate resources for all application microservices within the same cluster. In that case, our federated clustering approach allows the framework to place the services across several multiple clusters. 
In that case, the root orchestrator iteratively requests the clusters in the priority list to search for a locally optimal worker for service deployment. 
%
In case of resource failures, which are 
highly probable at the edge, the service manager of the associated cluster marks all affected services as failed. 
The orchestrator then attempts to re-deploy each service on another resource that satisfies the SLA requirements within the same cluster.
If unsuccessful, the rescheduling request is recursively propagated to the root orchestrator until a suitable worker is found~\cite{edisco}. 
Similarly, if the cluster orchestrator observes SLA violations of a running service, it triggers a migration.
Service migration follows a similar procedure as service rescheduling in case of failures, except the original service is terminated once the newer instance becomes operational.

Our delegated service scheduling approach significantly reduces the problem search space of multi-objective task placement at the edge by only considering a subset of candidate resources (limited to each cluster).
Note that in addition to our proposed ROM and LDP approaches, it is also possible to integrate extensive research on edge service scheduling~\cite{salaht2020overview} into our framework, as long as the algorithm can be re-architected to operate on a \emph{t-tier} hierarchy. 
However, we admit that our hierarchical scheduling mechanism sacrifices global optimality for reduced complexity.
%
Therefore, in the future, we plan to leverage recent investigations on this topic and incorporate heuristics and deadline-guarantee-based approaches~\cite{zhu2018task} to discover a near-optimal solution with increased probability.
%


\section{Service Communication}
\label{section:communication}

Supporting reliable intra-service (and service-to-user) networking in edge infrastructures can be quite challenging. 
Unlike cloud environments, edge infrastructures are susceptible to dynamic changes and hardware failures.
Application deployment is more fluid at the edge as services migrate to remain close to mobile users and dependent microservices~\cite{yi2017lavea}.
Furthermore, developers are likely to deploy several instances of their application microservices to cover a larger geographical area.    
As a result, traditional network load balancing techniques~\cite{loadbalancing} do not function well at the edge as clients are not only interested in connecting to the least loaded instance but also the one deployed \enquote{closest} to them.
Moreover, with multiple participating infrastructure operators, it is impractical to presume that every edge server will be accessible over a public network -- which is an implicit assumption for most existing orchestration frameworks~\cite{k3s:online,Kubernetes:online,mk8s:online}. 

\begin{figure}[!t]
    \centering
	\includegraphics[width=0.9\linewidth]{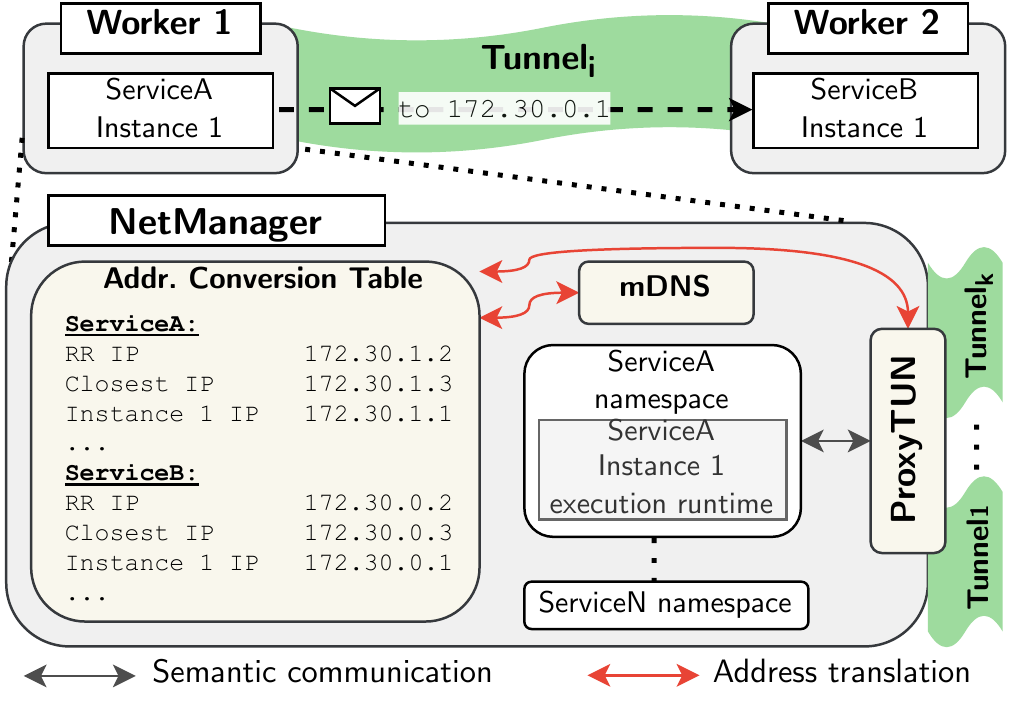}
	\caption{Service communication across edge servers.}
	\label{fig:netmanager}
	\vspace*{-1em}
\end{figure}
To overcome the challenges above,
we design a robust networking component (named \texttt{NetManager}) that (i) accommodates dynamic infrastructure changes and  (ii) transparently integrates resources from multiple operators without imposing overheads on application providers.
As shown in \Cref{fig:edgeio-schema}, \texttt{NetManager} is installed only at the workers, which allows us to separate the bulk of data-plane networking complexity from the control-plane operations. 
%
\Cref{fig:netmanager} shows the cross-section of the \texttt{NetManager} component enabling communication between two services (service A and B) deployed on different edge servers (worker 1 and 2).
We use logical IP addresses to decouple the physical address of the edge server from the address of the deployed service, thereby forming a service overlay that remains oblivious to underlying infrastructure changes. 
On top of that, the addressing mechanism binds multiple semantic IP addresses, namely \texttt{serviceIP}s, to each service operational on the worker. 
Each \texttt{serviceIP} maps to a different instance of that service in the infrastructure according to a balancing policy, which is tracked in the address conversion table (analogous to a routing table).
A \texttt{serviceIP}, drawing inspiration from semantic routing~\cite{king-irtf-semantic-routing-survey-03}, can be used by an application to find the instance that best suits that policy automatically.
For example, in \cref{fig:netmanager}, worker 1 maintains ServiceA's logical address for each instance as well as the \textit{closest} and \textit{round robin} \texttt{serviceIP}s. 

The address table also tracks \texttt{serviceIP}s of target services (in this case, ServiceB) which allows ServiceA to communicate with the closest instance of ServiceB using \textit{closest} \texttt{serviceIP}. 
The \texttt{NetManager} also includes local mDNS which enables services to use load balancing naming schemes instead of IP addresses (e.g. \textit{serviceB.closest}).
At the time $t=0$, the worker sets all entries in the conversion table, except the local service instance address, to \texttt{null}.
%
Suppose an operational service sends a network packet towards an unknown address. In that case, the node requests an IP resolution to the cluster orchestrator's service manager (see \circled{10} in \cref{fig:edgeio-schema}) and populates its table entries. 
%
Similarly, if the table data is insufficient to instantiate a connection or the \texttt{serviceIP} gives a network error, the worker explicitly requests its orchestrator for a conversion table refresh -- which is recursively propagated up the hierarchy until resolved.
Any future updates to the requested \texttt{serviceIP}s are automatically pushed to the worker by the orchestrator.
%

%
%
%
%
%

The \texttt{NetManager} natively supports end-to-end encrypted transport-layer tunneling, using \texttt{proxyTUN} to (i) transparently maintain the service overlay network across multiple nodes and (ii) allow safe traversal over untrusted networks.
%
%
Each service requests the \texttt{proxyTUN} to establish a connection to a \texttt{serviceIP}.
%
The \texttt{proxyTUN} first chooses the service instance based on the requested balancing policy, then translates the semantic address to the logical address through a table lookup.
%
%
The \texttt{proxyTUN} actively maintains and dynamically adjusts endpoints of the tunneled connections to adapt to infrastructure changes, 
meanwhile ensuring that the services continue to communicate uninterrupted.
%
For ingress/egress traffic outside the edge infrastructure (e.g., towards private end-users or third-party endpoints), the \texttt{proxyTUN} uses 
the service manager in cluster orchestrator as a VPN server that either redirects or tunnels the traffic on behalf of the worker.
In the future, we plan to utilize multipath transmissions using MPTCP~\cite{mptcpEdge, longitudinalMPTCP, mptcpAdoption} over multiple available network interfaces simultaneously at the edge to improve network reliability and availability~\cite{qaware}.

As the number of simultaneous service connections escalates, maintaining an increasing number of tunnels may become a burden for \texttt{proxyTUN}.
%
To overcome this, we distinguish between \textit{configured} and \textit{active} links. 
We consider a tunnel to be \textit{active} only when services are using it for data transmissions. 
%
In contrast, if a tunnel has been inactive for a while, e.g., the service using it has migrated, it is marked as \textit{configured} and becomes a candidate for garbage collection. 
%
%
%
Therefore, each resource $R^i_j$ has maximum $n-1$ outbound \textit{configured} links, one for every worker in the infrastructure, where $n=\sum_i(|C_i|-1)$ (excluding $R^i_{CO}$).
We can define $L$ as the set of all the \textit{configured} links 
$(R^i_l,R^j_m) \in L$ such that $ \forall {C_i,C_j \in C}; {0 \leq l < |C_i|}; {0 \leq m < |C_j|}$ and $R^i_l \neq R^j_m$.
%
%
Consequently, defining $k$ as the maximum number of \textit{active} tunnels that can be maintained in each worker node,  the set $A$ of the \textit{active} links is $A \subset L \;\; iff \;\; k<n\ $.
When the number of required tunnels exceeds $k$, the tunnel eviction mechanism uses the least recently used ($LRU$) policy. 
\section{Implementation: \sys}
\label{section:implementation}
%

As discussed in \cref{section:system_overview}, edge infrastructures can be composed of many heterogeneous devices with diverse form factors, hardware characteristics, and energy constraints~\cite{raspberry:online, intelnuc:online, coral:online}.
From a software perspective, these devices can have different runtime characteristics, including CPU architectures, virtualization support, etc.
%
%
An ideal orchestration platform must be capable of absorbing the inherent infrastructure heterogeneity without amplifying operational overheads -- all the while providing developers familiar techniques to seamlessly extend their cloud-supported applications to use the edge.
%
%

We implement our orchestration framework and its components (shown in \cref{fig:edgeio-schema}) as \sys.
%
Our comprehensive implementation, spanning over 11,000 lines of code, supports popular development tools and virtualization techniques necessary for edge computing.
We currently restrict \sys only to support a two-tier hierarchy (i.e., without sub-clusters) since the topology already embraces most edge computing models~\cite{noghabi2020emerging, promise_of_edge} and can sufficiently demonstrate the benefits of hierarchical orchestration at the edge.
However, \sys can be extended to support multiple hierarchy tiers with relative ease.
%


\smallskip
\noindent \textbf{Orchestration.}
We implement the root and cluster orchestrators in $\approx$ 4500 and 2800 lines of Python code, respectively. 
Infrastructure operators can initialize a cluster of edge servers as workers attached to a cluster orchestrator and register it with the root. 
The key technical difference between the two orchestrators is the communication protocol for the control plane traffic. 
%
Within each cluster, control message exchange between workers and the orchestrator (i.e., worker statistics, scheduling directives, etc.) is over MQTT, which is a lightweight message-passing networking protocol that allows \sys to scale cluster sizes without communication overheads efficiently.
On the other hand, the interaction between cluster and root orchestrator utilizes HTTP(S) WebSockets, which implicitly allows us to monitor the liveness of both orchestrator endpoints and trigger remedial actions in case of failures.
%




\smallskip
\noindent \textbf{Service Scheduling and Management.} 
Application providers can deploy their services at the edge by submitting the SLA and code binaries to the root orchestrator.
Using our delegated scheduling scheme, the scheduler component of the root and cluster orchestrator finds a suitable placement for the services in the infrastructure.
%
We implement both ROM and LDP service scheduling algorithms (discussed in \cref{section:scheduling}) as \emph{language-agnostic} \textit{plugins} to \sys's scheduler.
Our purposeful design choice allows future extensions as researchers/developers can easily incorporate custom scheduling algorithms without significant implementation overhead. 
%
Furthermore, since the scheduling logic of each tier is independent and decoupled, cluster operators can fine-tune the service scheduling and resource utilization behavior of their cluster by customizing the algorithm to their preference.
%

\sys also keeps track of deployed applications throughout their lifecycle 
through a state machine.
Each service instance starts as \textbf{\textit{requested}}, indicating that the root scheduler has initiated the scheduling process.
%
%
Once the cluster orchestrator finds a suitable worker for the service deployment, the service state becomes \textbf{\textit{scheduled}}. 
The worker deploys the service instance and periodically reports the current QoS and current resource utilization ($\cup(A^i)$) to the cluster orchestrator -- changing the service status to \textbf{\textit{running}}. 
As discussed in \cref{section:scheduling}, if the cluster orchestrator observes lapses in expected service behavior, e.g., the service becomes unavailable or violates its SLA, it triggers an implicit migration request
%
that is handled as a (new) scheduling request.
%
%
Once the migrated service instance becomes operational, the previous instance is undeployed, and its status changes to \textbf{\textit{terminated}}.
\sys also supports service \emph{replication} which follows a similar procedure as service migration, except for the termination of original service.
%
%
%
%
%
%
In case of unexpected early termination or failures, the affected service instances are marked as \textbf{\textit{failed}}.






\smallskip
\noindent \textbf{Networking.} 
We implement the \texttt{NetManager} component (see \cref{fig:netmanager}) of NodeEngine in $\approx$ 3,000 lines of Go code to ensure a low resource footprint without sacrificing performance.
%
%
The worker nodes obtain a unique subnetwork upon registering with their cluster orchestrator during the initial handshake.
Each deployed service is mapped to a logical address in the local subnetwork.
The \texttt{NetManager} creates a bridge that connects the virtual interface of the service instance to the \texttt{ProxyTUN} component that is responsible for ingress/egress traffic.
The \texttt{ProxyTUN} handles 
\texttt{ServiceIP} resolution (e.g., round-robin, closest) in a separate thread and finally establishes UDP-based connection tunnels to the resolved node/interface.
%
%
Additionally, the \texttt{NetManager} utilizes the MQTT communication channel between the worker and the cluster orchestrator for receiving service routing updates, e.g., in case of scaling, migration and undeployment.
\section{Evaluation}
\label{section:evaluation}









In this section, we evaluate and compare \sys to state-of-the-art orchestration frameworks through a series of experiments in realistic infrastructure testbeds (\cref{subsection:systemevaluation}). 
We also investigate the effectiveness of our service scheduling solutions (\cref{eval:servicescheduling})
and showcase \sys's ability to support the operation of realistic edge application workloads (\cref{eval:pipeline}). 


\subsection{Experiment Setup}
\label{subsection:experimentalsetup}
\begin{figure}[t]
    \centering
    \includegraphics[width=\linewidth]{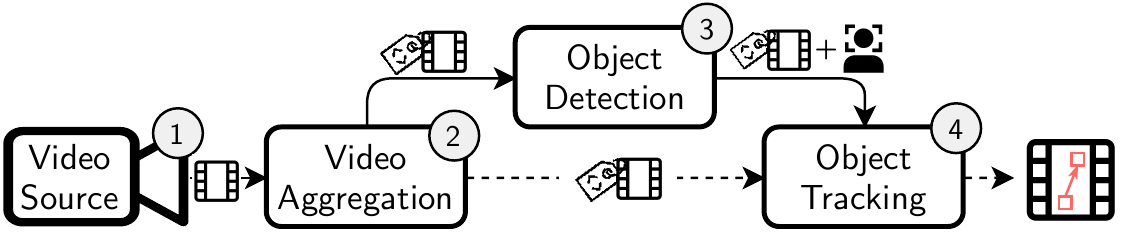}
    \caption{Cross-section of video analytics application.}
    \label{fig:pipeline}
    \vspace{-1em}
\end{figure}
%
%
%
We set up two different infrastructure testbeds for our evaluation.
Our \emph{High-Performance Computing (HPC)} testbed is a large VM-based compute cluster which allows us to configure different experiment configurations in a controlled environment.
We use VMs of different sizes for our tests, namely \emph{S}, \emph{M}, \emph{L}, \emph{XL} with 1, 2, 4, 8 GB RAM and 1, 2, 4, 8 CPUs, respectively. 
All VMs used Ubuntu 18.04 LTS over x86 processors.
Our other testbed is a \emph{Heterogeneous (HET)} edge-like cluster composed of devices with different hardware configurations, e.g., Raspberry Pis~\cite{raspberry:online}, Intel NUCs~\cite{intelnuc:online}, mini-desktops, and Nvidia Jetson AGX Xavier~\cite{AI_nvidia_jetson-agx_usecases:online}.
While VMs in our HPC testbed run on servers interconnected by 1 Gbps ethernet, our HET devices are interconnected via a mix of WiFi and ethernet links.
%
As a result, our \emph{HET} setup closely emulates a realistic deployment of edge computing with wirelessly-connected constrained resources.  

%
%
%
%
We attempted to compare \sys's performance against the most popular orchestration frameworks for our system evaluation.
However, despite our careful management, KubeFed~\cite{kubefed}, KubeEdge~\cite{kubeedge}, ioFog~\cite{Eclipse_iofog:online} and Fog05~\cite{Eclipse_fog05:online} performed quite inconsistently in our setup, often exhibiting random failures.
We attribute this behavior to their relatively early development stage and omit them from our analysis. 
As a result, we compare \sys against Kubernetes (K8s)~\cite{Kubernetes:online}, MicroK8s~\cite{mk8s:online} and K3s~\cite{k3s:online}, all of which are widely used real-world production-ready systems and have been proposed to operate at the edge~\cite{bohm2021profiling, rearchitecting-kubernetes-for-the-edge}. 
%
%

\begin{figure*}[t!]
    \begin{subfigure}[b]{0.30\linewidth}
        \includegraphics[width=\linewidth]{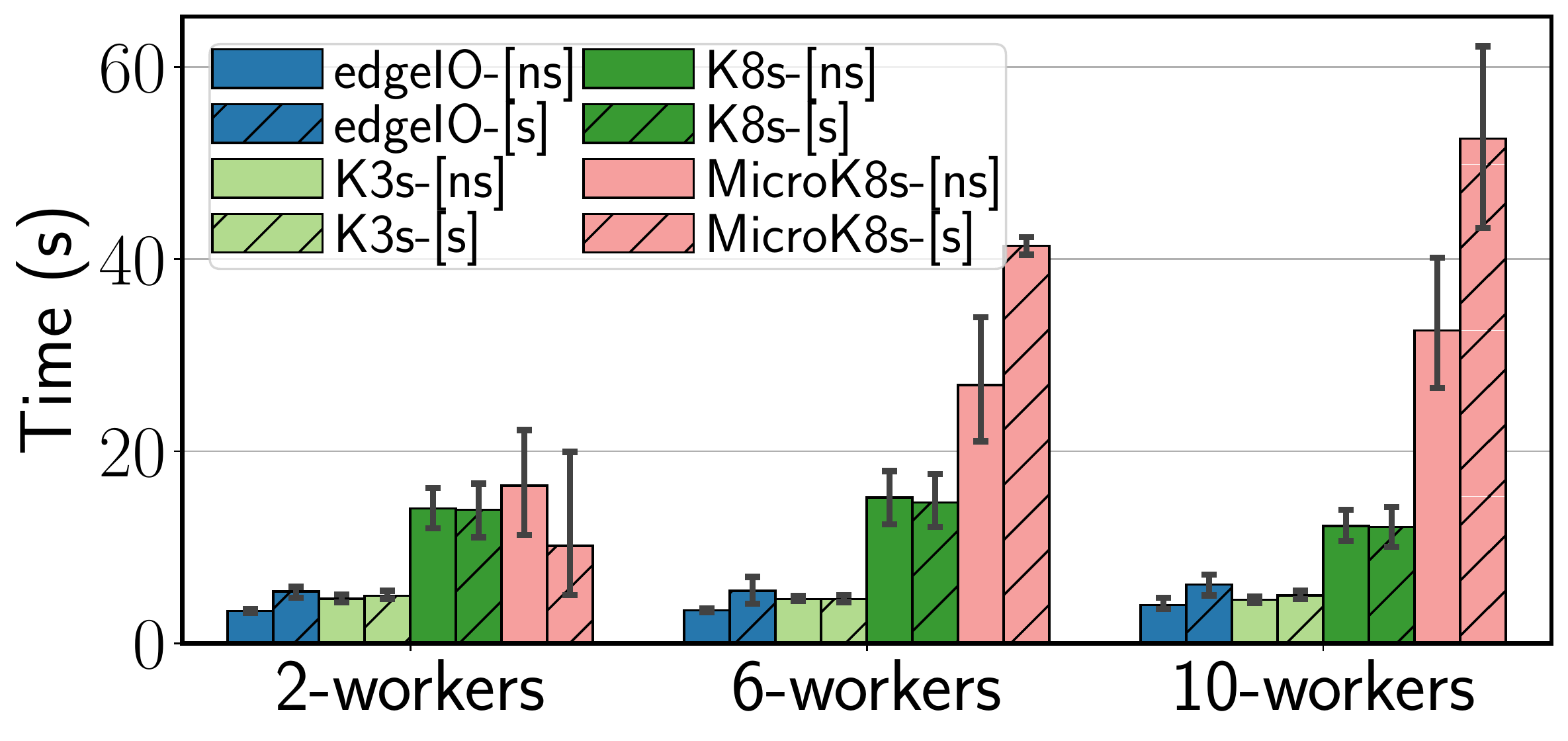}
        \caption{Deployment time}
        \label{fig:deployment-time}
    \end{subfigure}
    \hfill
    \begin{subfigure}[b]{0.34\linewidth}
        \includegraphics[width=\linewidth]{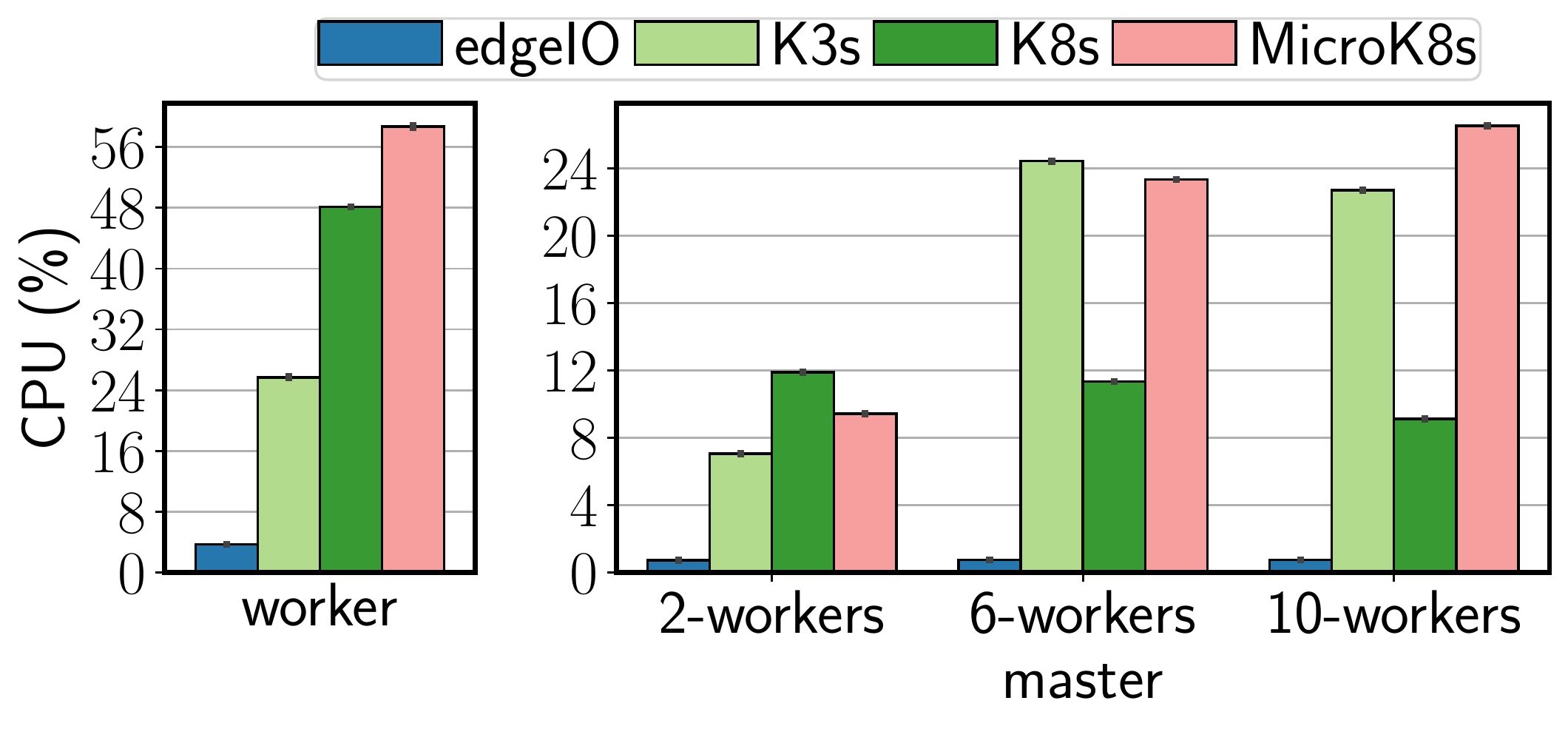}
        \caption{CPU utilization}
        \label{fig:cpu-utilization}
    \end{subfigure}
    \hfill
        \begin{subfigure}[b]{0.34\linewidth}
        \includegraphics[width=\linewidth]{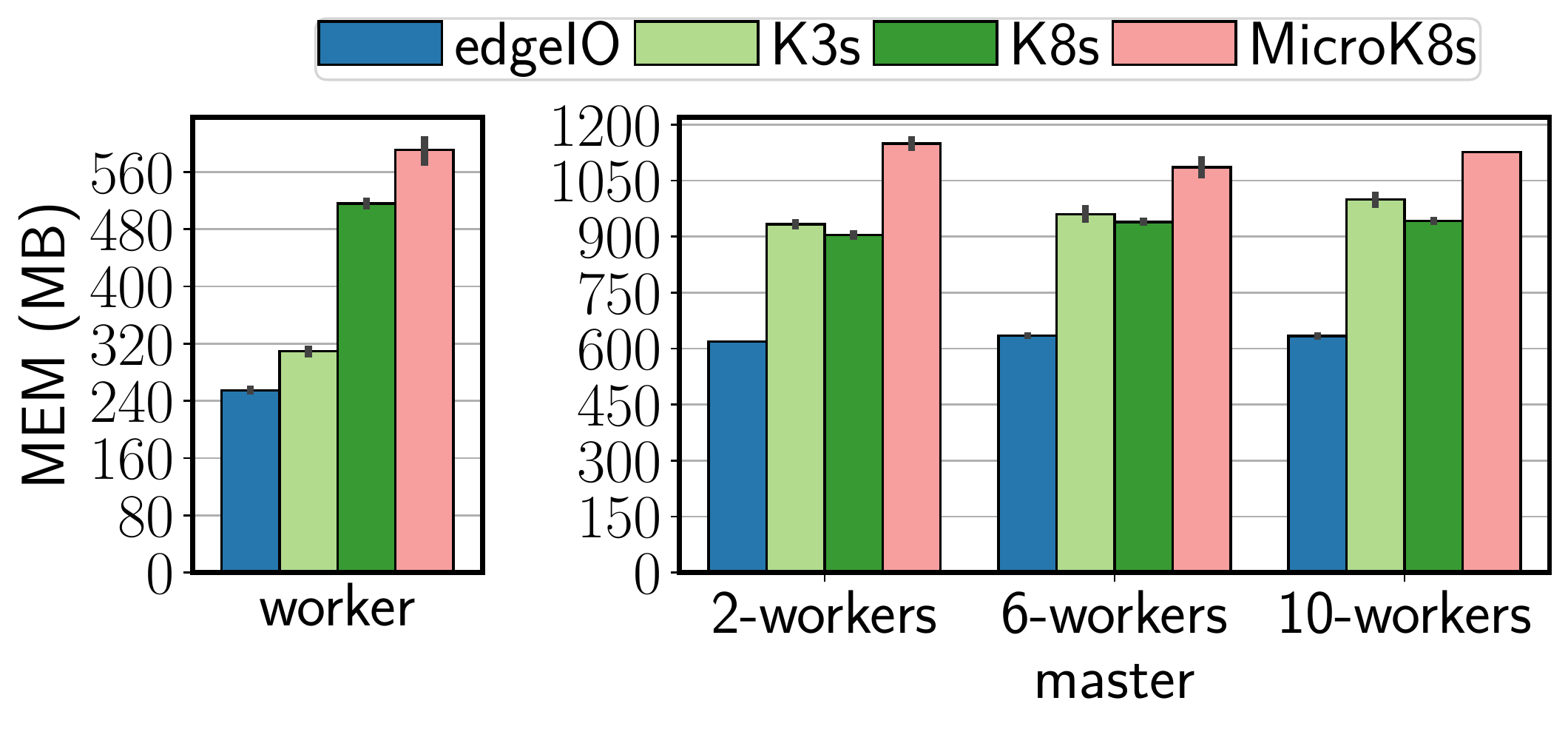}
        \caption{Memory utilization}
        \label{fig:mem-utilization}
    \end{subfigure}
    \caption{Orchestration framework performance for different infrastructure sizes.}
    \label{fig:system-load}
\end{figure*}
\begin{figure}[t!]
    \centering
    \begin{minipage}{0.485\linewidth}
        \centering
        \includegraphics[width=0.87\linewidth]{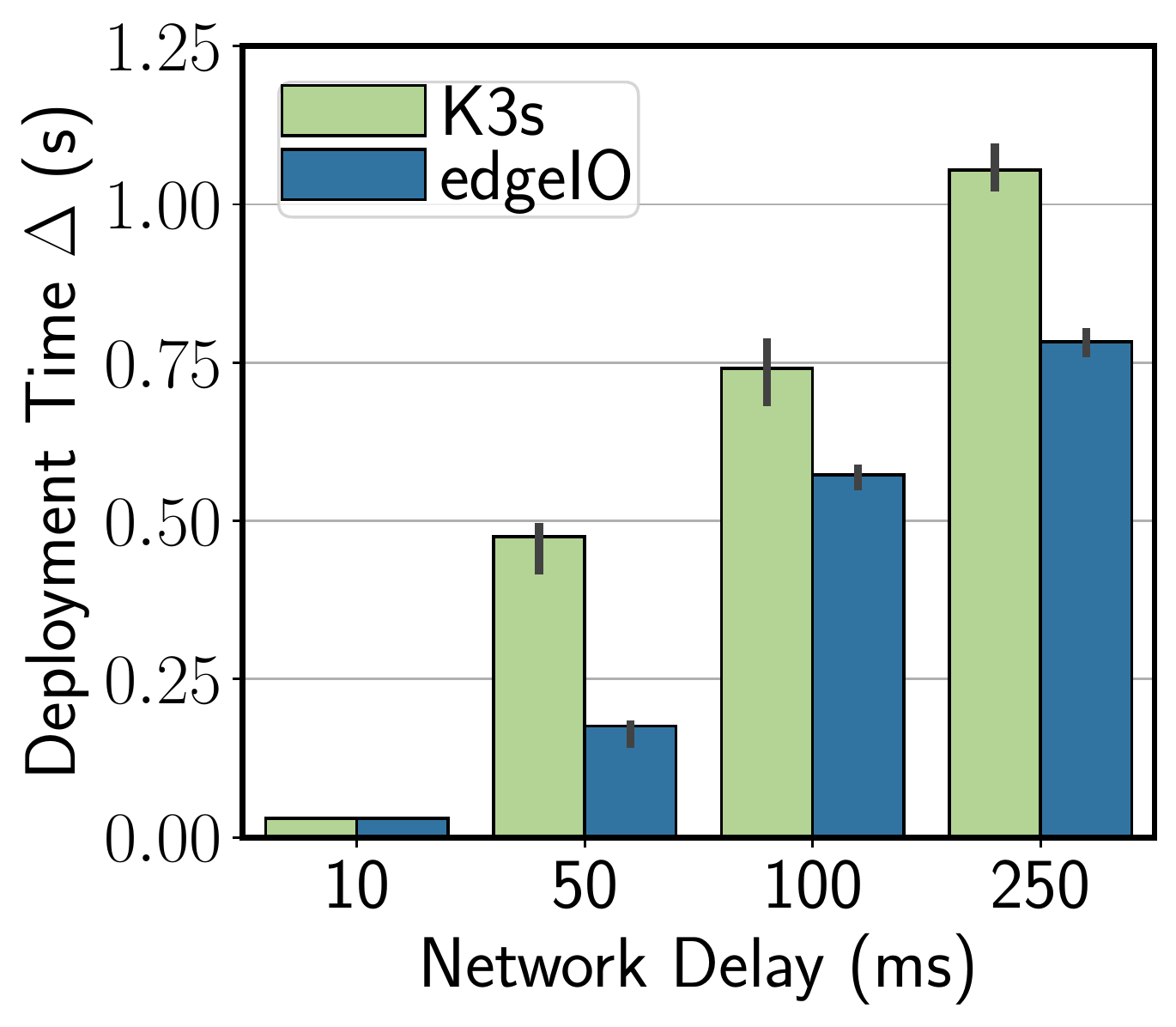}
         \caption{Deployment time with increasing network delay.}
        \label{fig:deployment-time-delay}
    \end{minipage}\hfill
        \begin{minipage}{0.475\linewidth}
        \centering
        \includegraphics[width=\linewidth]{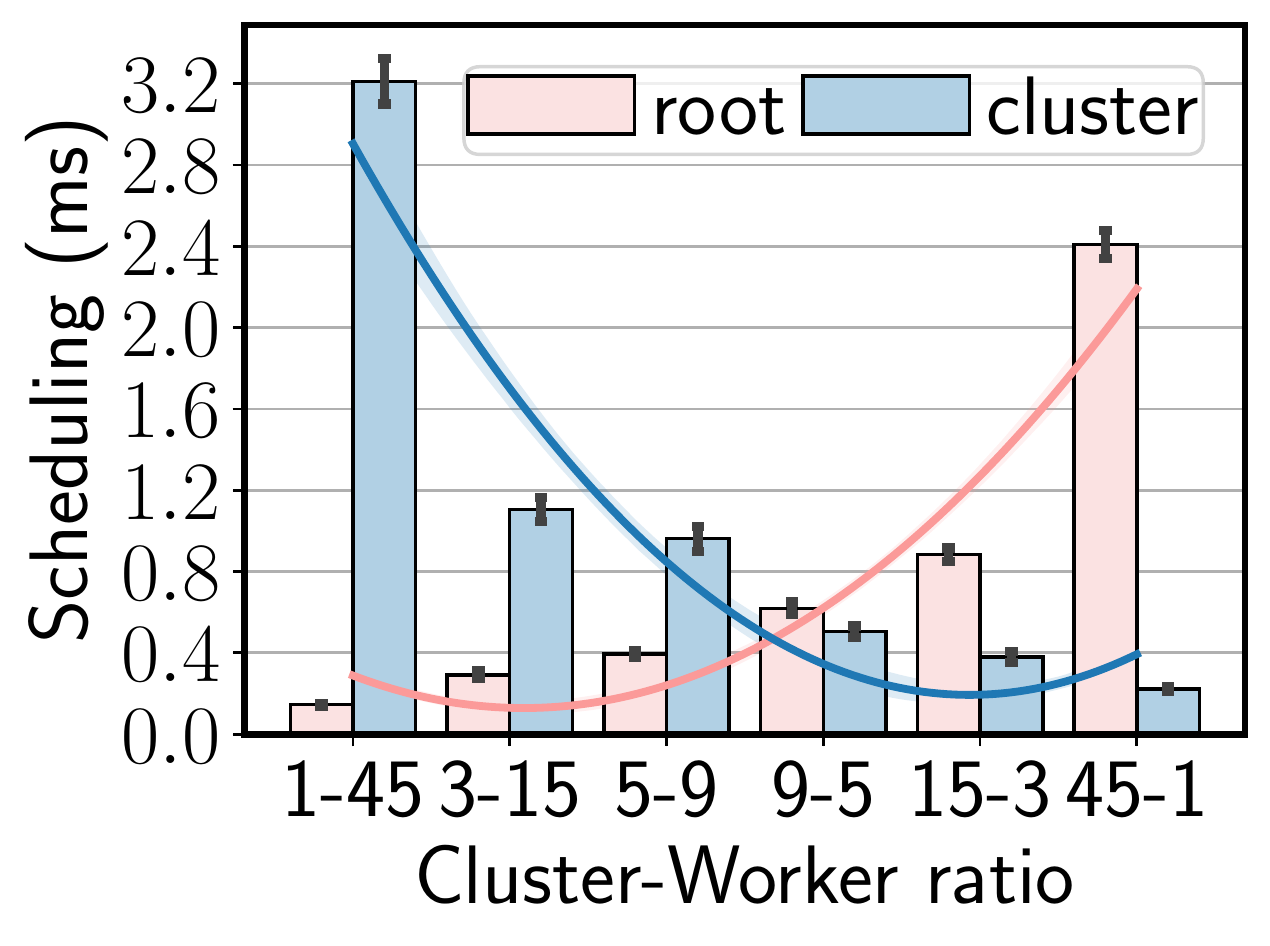}
         \caption{Schedule processing for cluster-worker ratios.}
        \label{fig:schedulingOverhead}
    \end{minipage}
    \vspace*{-1.5em}
\end{figure}

We use two different application workloads for our evaluation.
For our stress-test experiments, we utilize an Nginx web server which allows us to control the operational load on workers dynamically. 
We also developed a (live) video analytics application that detects and tracks objects in a video, which has been touted as the killer application for edge computing~\cite{ananthanarayanan2017real}. 
The pipeline (shown in \cref{fig:pipeline}) is composed of four microservices~\cite{comb}. 
The \textbf{video source} sends an RTP encoded video stream \circled{1} and can be replaced with a live camera.
For repeatable experiments, we use the wildtrack dataset~\cite{wildtrack} as our source.
%
%
%
%
The \textbf{video aggregation service} \circled{2} stitches multiple camera feeds together and performs some pre-processing for the rest of the pipeline.
The \textbf{object detection service} \circled{3} uses YOLOv3 to detect objects in every frame. 
The processed metadata is sent to the \textbf{object tracking service} \circled{4}, which tracks the movement of each detected object across frames.
All application services were virtualized as Docker containers.
We repeat all our experiments \emph{at least} ten times across multiple days.
%
%
Only one framework was operational at any given time, and we flush the memory and disk of all resources at the end of each run to avoid artifacts due to residual files.
%
%
%
%
Unless otherwise specified, we consolidate all workers in \sys within a single cluster to architecturally resemble and remain comparable to other master-slave orchestration platforms (with cluster orchestrator analogous to master).

\subsection{Orchestration Performance}
\label{subsection:systemevaluation}
%

%
%
%
\smallskip
\noindent \textbf{Service Deployment.}
\Cref{fig:deployment-time} compares the time taken by each framework to deploy a low-footprint containerized Python application that tracks its deployment time. 
To emulate a constrained edge environment, we configure \emph{XL} VM as root, \emph{L} VM as cluster orchestrator in \sys and master in other platforms. 
All platforms use \emph{S} VMs as workers. 
%
We increase cluster size from 2 to 10 workers and measure overheads due to the service scheduler of each framework by toggling its operation, shown with \emph{s} (scheduler) and \emph{ns} (no scheduler). 
%
%
We observe that MicroK8s and Kubernetes (K8s) perform significantly worse ($\approx$ 10$\times$ slower for MicroK8s) than \sys.
MicroK8s' performance degrades considerably with increasing infrastructure size. 
We also find that for frameworks other than MicroK8s, scheduler operation adds almost negligible overhead to service deployment.
Our results are in line with other recent measurements on the topic~\cite{bohm2021profiling}.
Note that \sys exhibits minimal service deployment time, which remains unaffected by infrastructure size.

Since K3s's performance closely matched \sys, we tested both platforms in our HET testbed and gradually degraded the network conditions by using \texttt{tc} utility. 
\Cref{fig:deployment-time-delay} shows that \sys's deployment time performance surpasses K3s by $\approx$ 20\% with increasing network delays.
We observe similar behavior with packet losses on the network as well.
Specifically, \sys consistently outperformed K3s, achieving $\approx$ 50\% and 60\% reduction in deployment time with 20\% and 50\% losses, respectively (plot not shown for brevity). 
%
%
%
%
%
%
Note that the above experiments use a single cluster configuration, which is the worst-case scenario for \sys's multi-cluster orchestration.
To better showcase the capabilities of \sys, we record the time taken by the root and cluster scheduler for a different number of clusters and workers per cluster configurations (see \cref{fig:schedulingOverhead}).
It can be observed that \sys achieves better performance when the workers are somewhat balanced across multiple clusters in the hierarchy (see minima around nine clusters with five workers setting).

\begin{figure}[t!]
    \begin{subfigure}[b]{0.31\linewidth}
        \includegraphics[width=\linewidth]{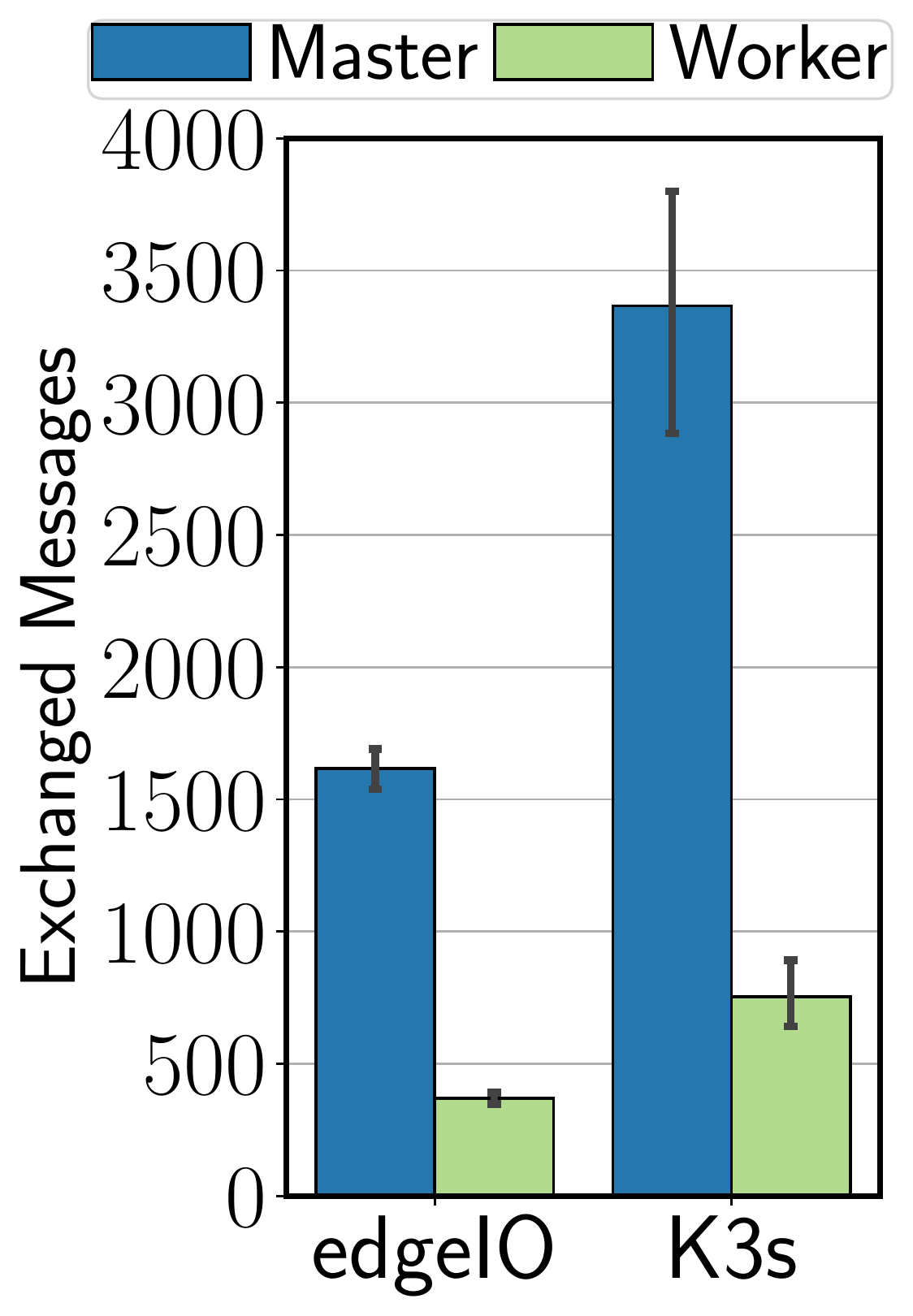}
        \vspace*{0.1em}
        \caption{Total control message overhead.}    \label{fig:control-messages}
    \end{subfigure}
    \hfill
    \begin{subfigure}[b]{0.69\linewidth}
        \includegraphics[width=\linewidth]{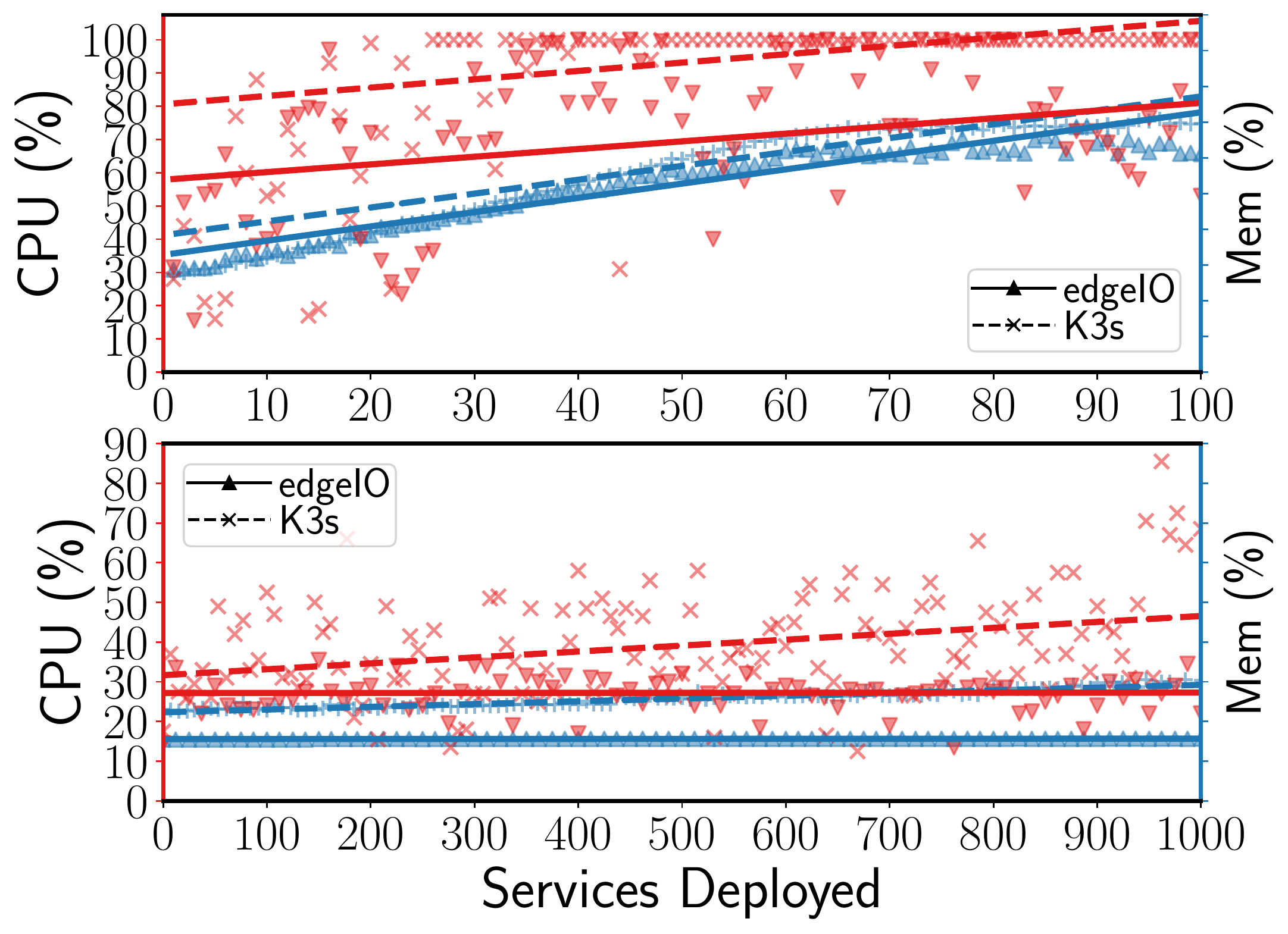}
        \caption{CPU (red)/memory (blue) usage of worker (top) \& cluster orch. (bottom) during stress test.}
        \label{fig:stress-test-cluster}
    \end{subfigure}
    \caption{Orchestration overhead in 10 node cluster. }
    \label{fig:stress-test}
    \vspace*{-1.5em}
\end{figure}

\smallskip
\noindent \textbf{Scalability.}
\Cref{fig:cpu-utilization,fig:mem-utilization} compares each orchestration platform's idle resource consumption in the HPC testbed with cluster size ranging from two to ten workers. 
%
%
Lower overhead at the worker indicates the platform's capability to operate on constrained devices.
On the other hand, lower overhead at the master highlights the platform's ability to handle scale.
%
%
We observe that all frameworks consume considerably more CPU for their operation than \sys.
%
%
Between the competitors, we find that K3s has a low worker node footprint while K8s supports scaling better as its performance in the master remains relatively consistent.
%
%
On the other hand, thanks to its lightweight design, \sys can support both scale and resource constraints at the edge as it achieves $\approx$ 6$\times$ reduction in CPU and $\approx$ 18\% memory usage on the workers along with $\approx$ 11$\times$ less CPU and $\approx$ 33\% less memory on the master. 
%
%
%
%

\begin{figure*}[th!]
    \begin{subfigure}[b]{0.50\linewidth}
         \includegraphics[width=\textwidth]{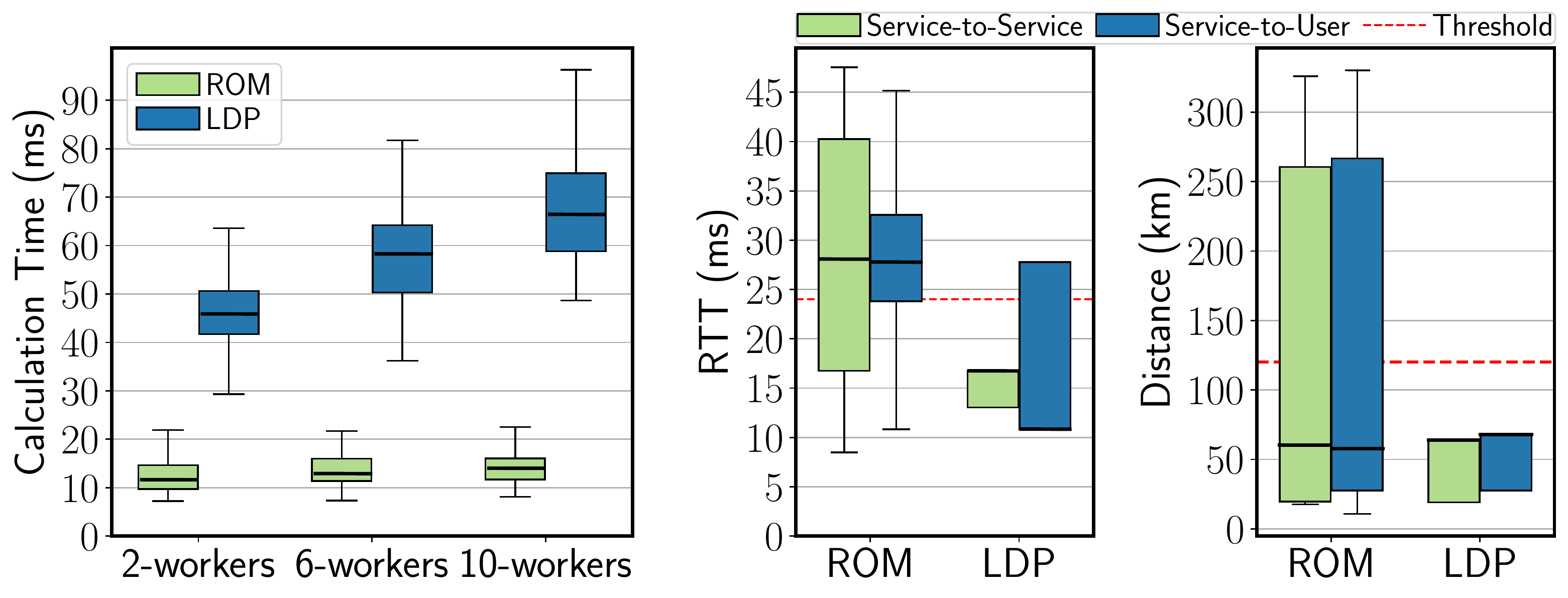}
         \caption{HPC testbed.}
        \label{fig:hpi-scheduling-results}
    \end{subfigure}
    \begin{subfigure}[b]{0.49\linewidth}
         \includegraphics[width=\textwidth]{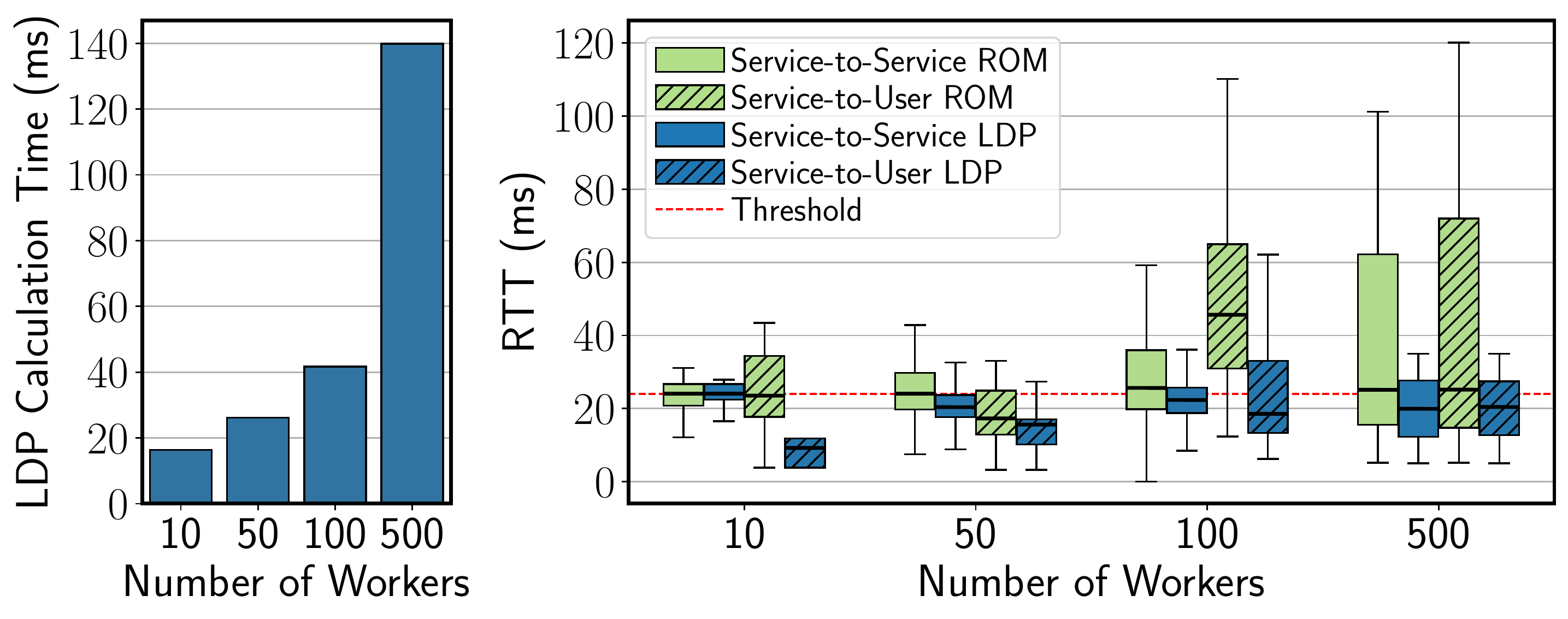}
         \caption{Simulated edge environments.}
        \label{fig:sim-scheduling-results}
    \end{subfigure}
    \caption{Resource-Only Match (ROM) and Latency and Distance Aware Placement (LDP) scheduler performance.}
    \label{fig:scheduling-results}
    \vspace*{-1.2em}
\end{figure*}

\begin{figure}[t]
    \begin{subfigure}[b]{0.535\linewidth}
        \includegraphics[width=\textwidth]{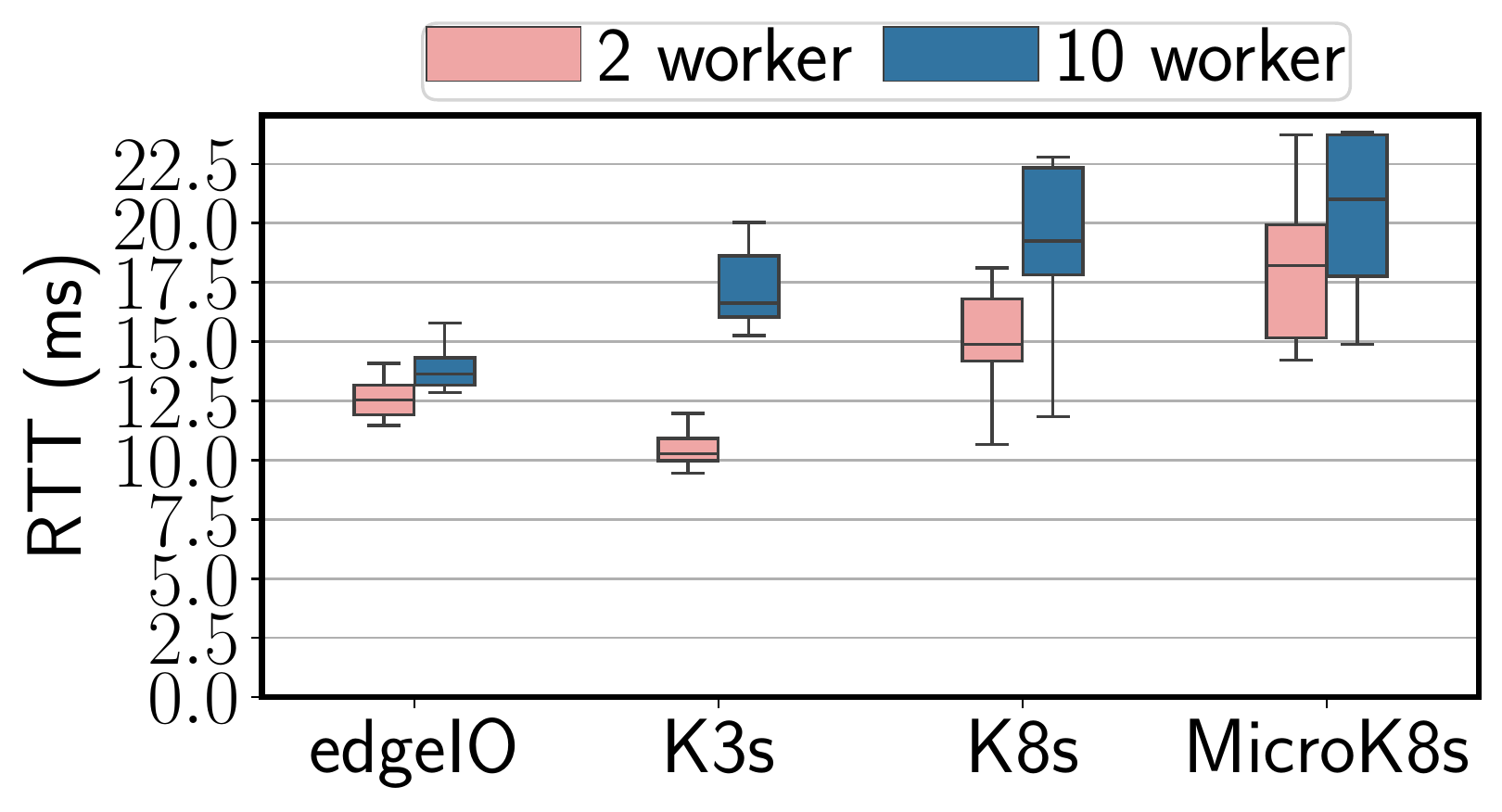}
        \vspace*{0.8em}
        \label{fig:network-overhead}
    \end{subfigure}
    \hfill
    \begin{subfigure}[b]{0.45\linewidth}
        \includegraphics[width=\textwidth]{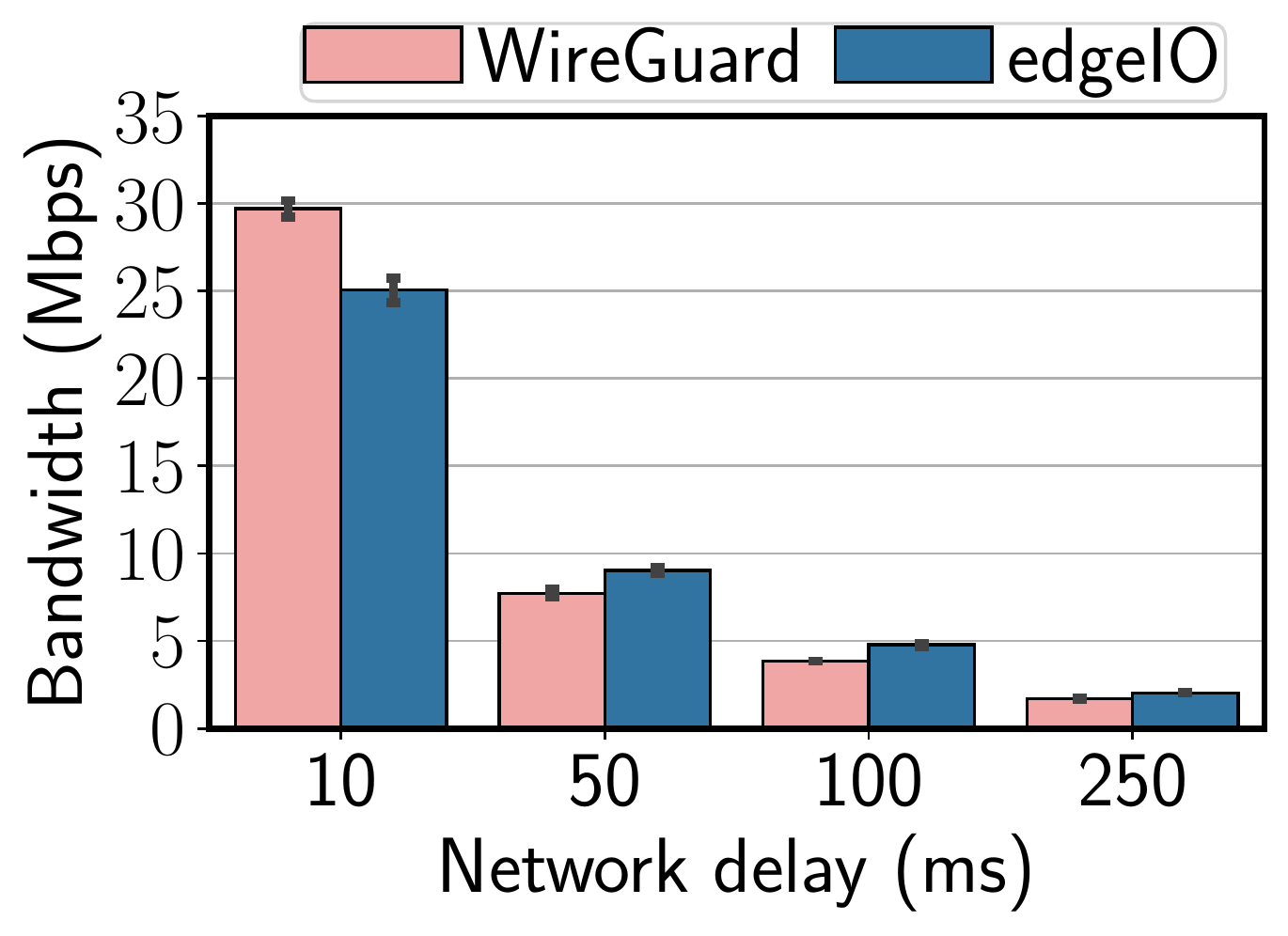}
        \label{fig:bandwidth}
    \end{subfigure}
    \vspace*{-10mm}
    \caption{Network latency and bandwidth overheads.}
    \label{fig:networking}
\end{figure}

\Cref{fig:stress-test} shows the CPU, memory, and bandwidth overhead of \sys against K3s for increasing service deployment. 
%
From \Cref{fig:control-messages}, we find that K3s sends $\approx$ 2$\times$ more control messages (from both worker and master) compared to \sys on the network for orchestration operations.
Not only does the increased messages indicate potential overhead of K3s in constrained edge network, but also highlights its dependency on network conditions for optimal operation --
justifying our results in \cref{fig:deployment-time-delay}. 
\Cref{fig:stress-test-cluster} compares the resource consumption in a cluster of 10 workers as we increasingly schedule up to 100 Nginx containers on each worker (totaling 1000 containers in the cluster).
%
The top half illustrates the worker's utilization, while the bottom shows the performance of the cluster orchestrator (or K3s master). 
Note that the primary overhead in the master is due to the scheduling operation.
We find that the \sys orchestration causes almost negligible overhead and can support numerous services effectively, achieving $\approx$ 10--20\% better performance than K3s. 
Similarly, the low footprint of \sys has significant operational advantages for the workers.
While K3s exhausted the available CPU at the worker at $\approx$ 60 services, \sys was able to deploy 100 services with 30\% CPU still available.
%
%

%

\smallskip
\noindent \textbf{Networking.}
We now compare the performance of \sys's networking component to the state-of-the-art.
For our experiments, we set up multiple Nginx server replicas on different workers, and then deploy a single client performing GET requests to a server instance.
%
%
\Cref{fig:networking} (left) shows the average round-trip latency between the client and the closest server.
A lower RTT indicates the effectiveness of load balancing by the platform.  
%
%
On average, K3s performs $\approx$ 10--20\% better than \sys in a single client-server setup, while Kubernetes and MicroK8s perform considerably worse -- likely due to their substantial operational overhead in constrained resources (as noted in \cref{fig:system-load}).
%
With multiple server replicas, the performance of all competitors degrades significantly, resulting in RTT inflation of $\approx$ 20\% compared to \sys.
%
Closer investigation revealed the cause of network overhead in a single service instance setup to \sys's traffic tunneling.
Recall from \cref{section:communication} that \sys uses L4-tunneled traffic to allow communication across cluster networks.
%
%
We evaluate the impact of \sys's tunneling on network performance and compare it with \texttt{WireGuard}~\cite{wireguad:online} -- an open-source tunneling solution used by most frameworks.
%
%
We emulate the network inconsistencies at the edge~\cite{mohan2020pruning} by gradually increasing the delay between the client and the servers from 10 to 250 ms.
%
%
\Cref{fig:networking} (right) compares the time to download a 100~MB sparse file over HTTP using both approaches.
%
%
While \texttt{WireGuard} achieves $\approx$ 10\% higher bandwidth than \sys in low latency settings, the performance gap diminishes with network delays.
%
%
We also measured the performance of both systems for different loss rates (1\% to 10\%) and always found \sys to be in competitive range (2--10\%) of \texttt{WireGuard}.
%
%
We must stress that \sys is in its early development stages. 
We plan to investigate alternate approaches such as L7 connection-level tunneling instead of current L4 per-packet tunnels in the future.

\subsection{Service Scheduling}
\label{eval:servicescheduling}

%
We now evaluate our proposed ROM and LDP schedulers in (a) \emph{HPC} with up to 10 workers and (b) simulated infrastructure with up to 500 edge servers.
%
While our HPC experiments provide us an insight into the real-world operation of our schedulers using \sys, our simulation experiments allow us to investigate the behavior of the schedulers at scale.  
%
We configure network latencies between edge servers within 10 - 250 ms, which, as per recent research~\cite{surrounded}, is a typical latency range between users and cloud datacenters globally.
%
We then instruct the schedulers (using SLA) to find workers that satisfy 1 CPU, 100 MB memory, $\approx$ 20 ms latency (usual for immersive edge applications~\cite{mohan2020pruning}) and 120 km operational distance. 
%
Figure~\ref{fig:hpi-scheduling-results} shows 
HPC results.
Since ROM only performs a best-fit match for overall computational requirements, its calculation time is significantly lower than LDP, whose computational complexity is much larger due to distance calculations and trilateration in the Vivaldi network.
However, LDP almost always satisfies the latency and geographical SLA constraints (shown as dashed red lines) albeit at a higher calculation cost which increases with infrastructure size.

%
We investigate the schedulers' behavior further in our simulation experiments (\cref{fig:sim-scheduling-results}) which shows the
LDP calculation time with up to 500 workers and achieved RTT latencies by both ROM and LDP. 
%
We find that LDP's calculation time escalates several manifolds with infrastructure size. 
However, the absolute time is still in the milliseconds' range, which may not be a significant overhead as the service is not yet operational.  
On the other hand, LDP can effectively support latency-based service constraints at the edge since it usually satisfies the latency thresholds even in large infrastructures (see RTT in 500 worker cluster).
%
We attribute minor lapses in latency thresholds to Vivaldi, 
whose accuracy is significantly affected by triangle inequality violations in large networks~\cite{vivaldi_ncs}.

\begin{figure}[t!]
    \begin{subfigure}[b]{0.32\linewidth}
        \includegraphics[width=\linewidth]{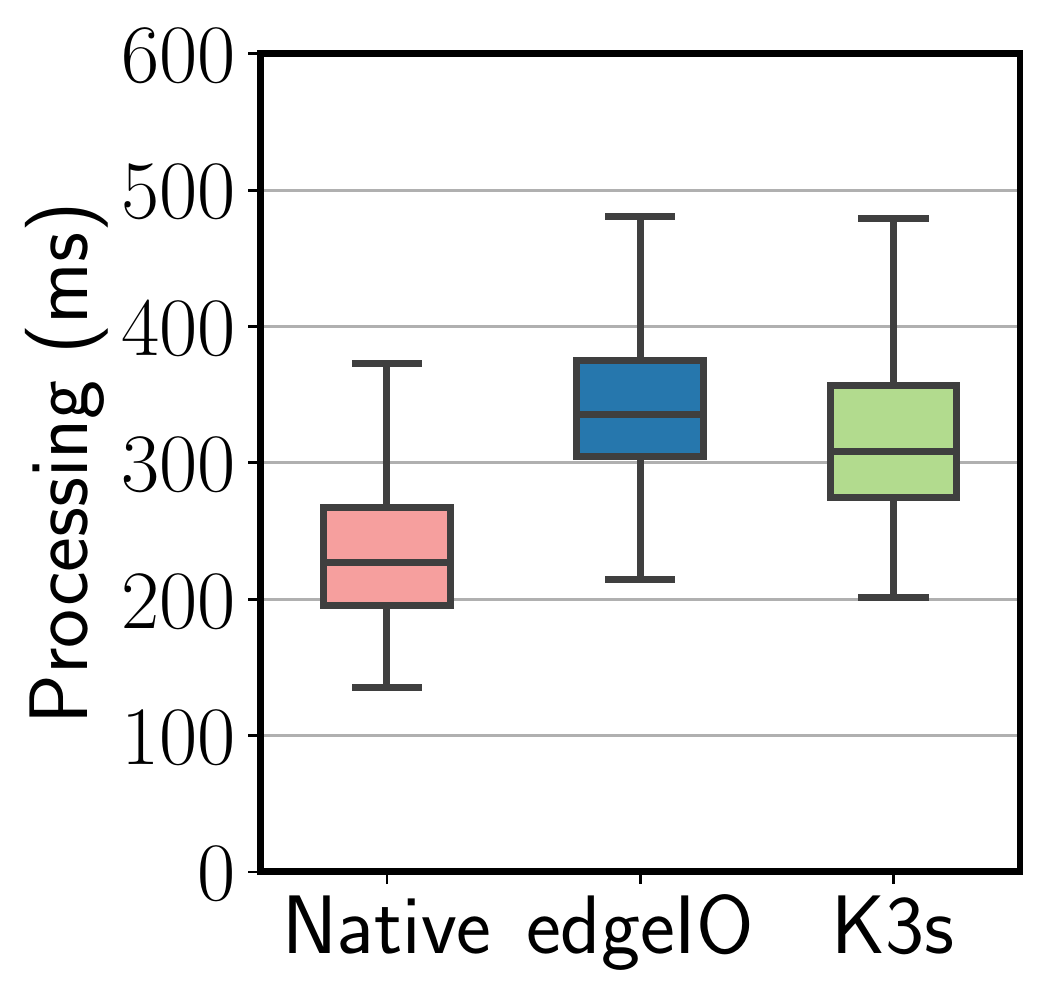}
        \caption{Tracking}
        \label{fig:tracking}
    \end{subfigure}
    \begin{subfigure}[b]{0.32\linewidth}
        \includegraphics[width=\linewidth]{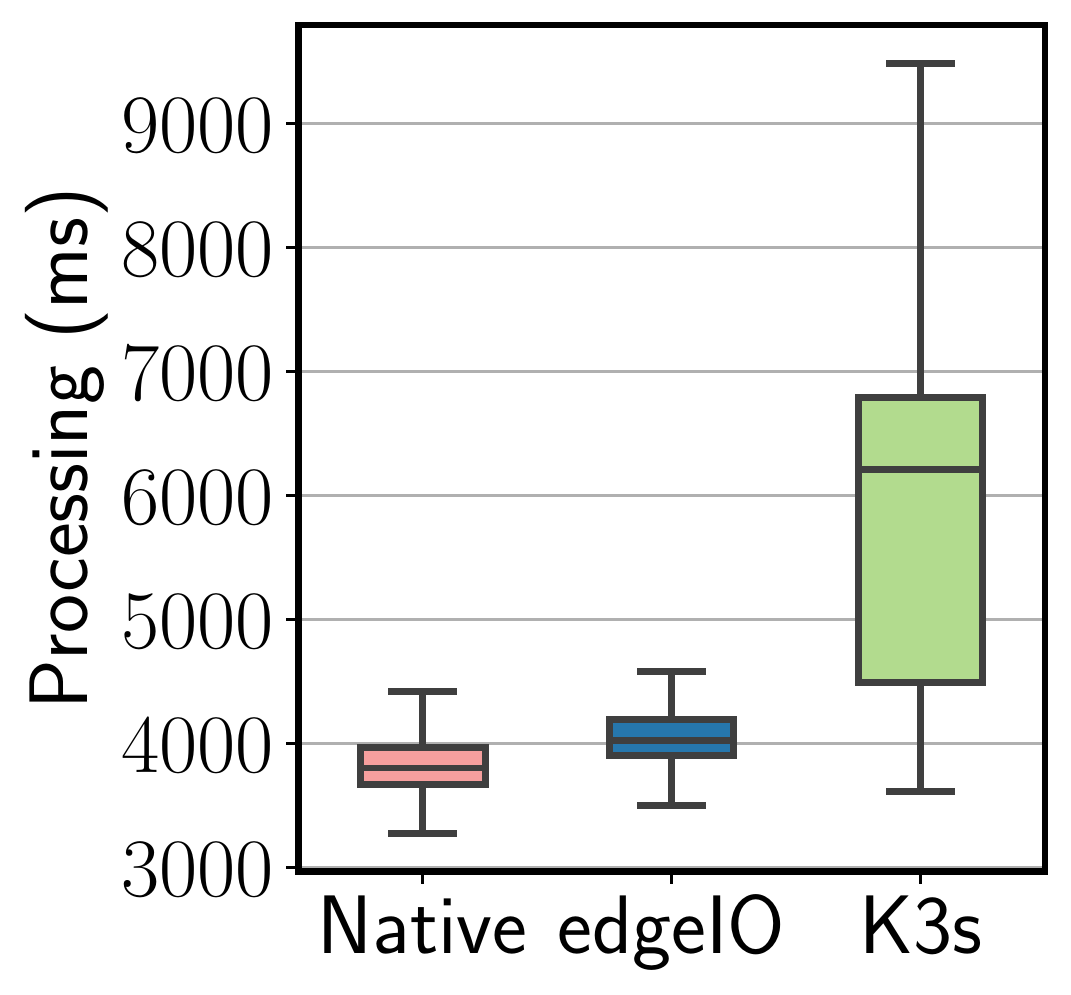}
        \caption{Detection}
        \label{fig:detection}
    \end{subfigure}
    \hfill
    \begin{subfigure}[b]{0.31\linewidth}
        \includegraphics[width=\linewidth]{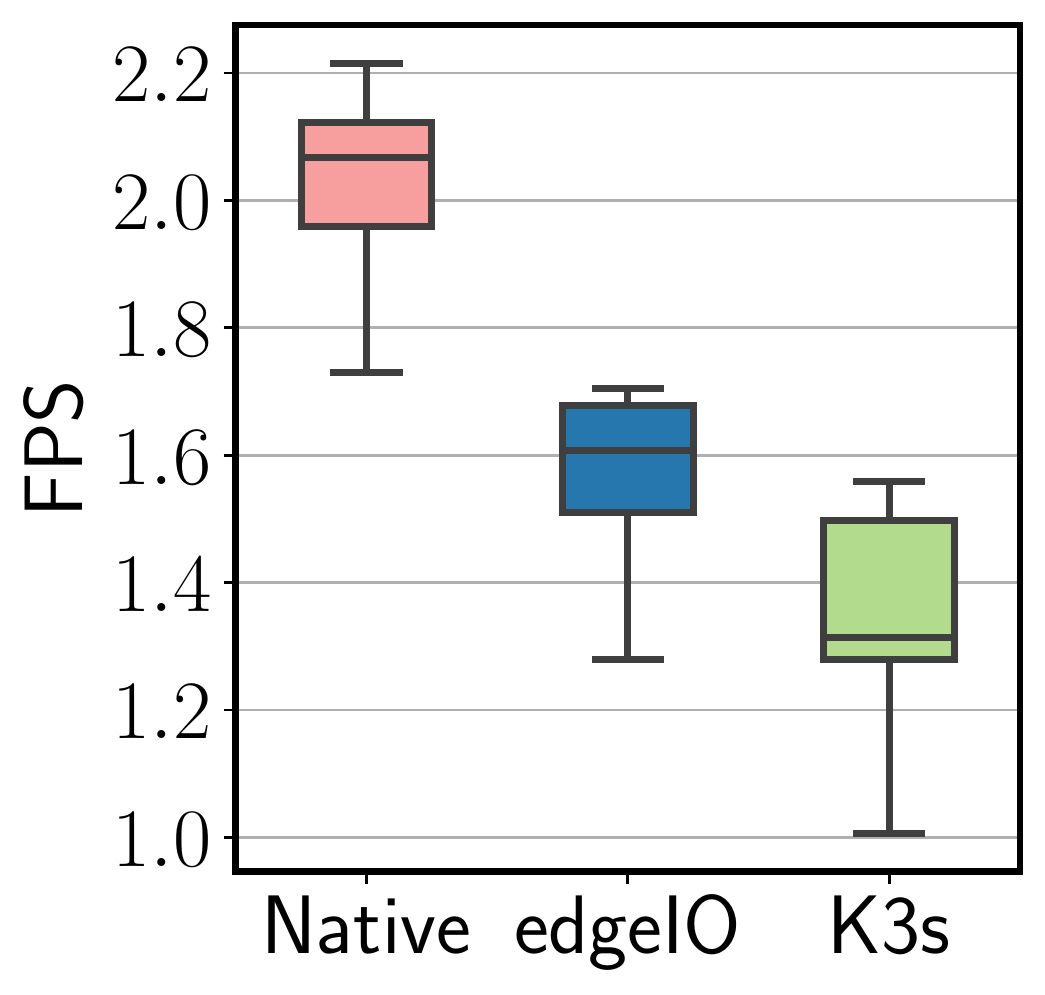}
        \caption{FPS}
        \label{fig:fps}
    \end{subfigure}
    \caption{Live video analytics application performance.}
    \label{fig:performance-metrics}
\end{figure}

\subsection{Realistic Edge Application Support}
\label{eval:pipeline}
%
%

%
We now investigate the capability of the orchestration frameworks to support the operation of the video analytics pipeline described in \cref{subsection:experimentalsetup}.
%
%
We create a cluster of four \emph{S} VMs as workers in our HPC testbed and map each microservice of the application to separate workers.
%
%
Interestingly, we observed that both Kubernetes and MicroK8s were unable to reliably support the application in our tests since their orchestration components consumed the majority of the resource's capacity (see \cref{fig:system-load}). 
As a result, we compare \sys's performance to K3s, and without any orchestration (named native) on the same infrastructure. 
Since the application can fully utilize the available resource capacity in the native setting, we consider it our baseline. 
%
%
\Cref{fig:performance-metrics} shows our results.
%
\sys and K3s exhibit similar performance for object tracking, taking $\approx$ 300-400 ms.
However, due to its minimal footprint \sys significantly outperforms K3s for supporting the more resource-demanding object detection, achieving results closer to the baseline.
%
Overall, application performance over \sys exceeded K3s by almost 10\%.
We also replicated our experiments in HET but skip its discussion as the application performance was similar to HPC.
%
%

\section*{Conclusion}

In this paper we presented a flexible orchestration framework designed to support diverse and heterogeneous edge computing environments.
Through our unique hierarchical cluster management, we allow multiple operators to participate in a federated infrastructure while retaining complete administrative control.
%
We also designed a delegated service scheduling mechanism and service overlay networks to effectively deploy and support application services in edge infrastructures spanning vast geographical regions over different networks.
%
%
%
%
%
We implemented \sys and thoroughly evaluated it against the state-of-the-art using various experiments in realistic edge testbeds.
\sys consistently outperformed its competitors, achieving $\approx$ 10$\times$ reduction in resource usage and 10\% improvement in application performance.
\balance

\bibliographystyle{IEEEtran}
\footnotesize
\bibliography{references}

\end{document}